\documentclass{aa}
\usepackage{graphicx}
\usepackage{natbib}
\usepackage{rotating}
\bibpunct{(}{)}{;}{a}{}{,}
\newcommand{\kms}{km~s$^{-1}$}
\newcommand{\brg}{Br$\gamma$}
\newcommand{\pad}{Pa$\delta$}
\newcommand{\hei}{\ion{He}{i}}
\newcommand{\htwo}{H$_{2}$}
\newcommand{\micron}{$\mu$m}
\newcommand{\mdot}{\ensuremath{\dot{M}}}

\begin{document}

   \title{VLT $K$-band spectroscopy of massive young stellar objects in
(ultra-)compact \ion{H}{ii} regions\thanks{Based on observations
collected at the European Southern Observatory at Paranal, Chile (ESO
programmes 64.H-0425, 65.H-0602, 68.C-0652 and 69.C-0448).}}

   \author{A.~Bik\inst{1,2} \and L.~Kaper \inst{1} \and 
           L.B.F.M.~Waters \inst{1,3}}

   \offprints{A.~Bik (abik@eso.org)}

   \institute{Astronomical Institute ``Anton Pannekoek'',
           University of Amsterdam,  Kruislaan 403, 1098 SJ Amsterdam,
           The Netherlands
      \and European Southern Observatory, Karl-Schwarzschild
           Strasse 2, Garching-bei-M\"unchen, D85748, Germany
      \and Instituut voor Sterrenkunde, Katholieke Universiteit Leuven,
           Celestijnenlaan 200B, B-3001 Heverlee, Belgium}

   \date{Received; accepted}

\authorrunning{A.\ Bik et al.}
\titlerunning{Massive YSOs in (ultra-)compact \ion{H}{ii} regions}

   \abstract{High-quality $K$-band spectra of strongly reddened point
sources, deeply embedded in (ultra-) compact \ion{H}{ii} regions, have
revealed a population of 20 young massive stars showing no
photospheric absorption lines, but sometimes strong \brg\
emission. The Br$\gamma$ equivalent widths occupy a wide range (from
about 1 to over 100~\AA); the line widths of 100--200~\kms\ indicate a
circumstellar rather than a nebular origin.  The $K$-band spectra
exhibit one or more features commonly associated with massive young stellar
objects (YSOs) surrounded by circumstellar material: a very red colour
$(J-K) \ga 2$, CO bandhead emission, hydrogen emission lines
(sometimes doubly peaked), and \ion{Fe}{ii} and/or \ion{Mg}{ii}
emission lines. The large number of objects in our sample allows a
more detailed definition and thorough investigation of the
properties of the massive YSOs.
In the ($K, J-K$) colour-magnitude diagram (CMD) the
massive YSO candidates are located in a region delimited by the OB
zero-age main sequence, Be stars, Herbig Ae and Be stars, and B[e]
supergiants. The massive YSO distribution in the CMD suggests that the
majority of the objects are of similar spectral type as the Herbig Be
stars, but some of them are young O stars. The spectral properties of
the observed objects do not correlate with the location in the
CMD. The CO emission must come from a relatively dense ($\sim 10^{10}
\mathrm{cm}^{-3}$) and hot ($T\sim 2000-5000$~K) region, sufficiently
shielded from the intense UV radiation field of the young massive
star. The hydrogen emission is produced in an ionised medium exposed
to UV radiation. The best geometrical solution is a dense and neutral
circumstellar disk causing the CO bandhead emission, and an ionised
upper layer where the hydrogen lines are produced. We present
arguments that the circumstellar disk is more likely a remnant of the
accretion process than the result of rapid rotation and mass loss such
as in Be/B[e] stars.  }

\maketitle

\section{Introduction}\label{sec:introduction}

The observational study of massive stars at the earliest evolutionary
phases is a rapidly expanding field. Although over the past decades
significant progress has been made in unravelling the formation process
of low-mass stars \citep[e.g.][]{Shu87}, it is not well understood how
massive stars form. The contraction timescale is short, so that
already very early in their formation massive stars will produce a
copious radiation field that may hamper or even reverse the accretion
process \citep{Wolfire87}. This has led to the suggestion that stars
more massive than $\sim 10 M_{\odot}$ cannot form through
(spherical) accretion alone, but instead form by collisions of
intermediate-mass stars \citep{Bonnell98}. Alternatively,
non-spherical accretion through a disk could solve the ``radiation
pressure problem'' \citep[e.g.][]{Yorke02}. Therefore, the detection
of circumstellar disks around young massive stars is regarded as an
essential step in understanding the formation of the most massive
stars. Observations at centimeter and millimeter wavelengths suggest
that rotating circumstellar disks are present around high-mass young
stellar objects \citep[e.g.][]{Minier98,Shepherd01Science,
Beltran04,Chini04,Jiang05,Patel05}.

Newly born massive stars are deeply embedded inside their natal
cloud, obscured from view by 10--100 magnitudes of visual
extinction. One of the first observable signatures is the radio and
infrared emission produced by the ultra- (or hyper-)compact \ion{H}{ii}
region that emerges when the contracting star becomes hot enough to
ionise the surrounding material. Although radio and infrared
observations provide information on the amount of ionising radiation,
these diagnostics do not allow an accurate determination of the
photospheric properties of the embedded young massive star(s)
\citep[for reviews, see e.g.][]{Churchwell91,Hanson98,Garay99,Churchwell02}.

Our strategy aims at the direct observation of the ionising star(s) of
(ultra-) compact \ion{H}{ii} (UCHII) regions at near-infrared
wavelengths, where the extinction is strongly reduced ($A_\mathrm{K}
\simeq 0.1 A_\mathrm{V}$), but the (reprocessed) emission by dust at
longer wavelengths has not yet set in. Recent developments in
near-infrared instrumentation have made this approach feasible.
By first identifying the candidate ionising stars in IRAS sources with
UCHII colours \citep[cf.][]{Kaper06} and subsequent $K$-band
spectroscopic follow-up \citep{Ostarspec05}, their properties can be determined.

 It turns out that our sensitivity allowed us to detect about  half of the UCHIIs in the near-infrared. As shown in \citet{Ostarspec05,Kaper06} and \citet{Hanson02}, the UCHII regions are always located in larger, extended HII regions. 
Apart from the near-infrared counterparts of some of the UCHII regions, we have spectroscopically identified  two other types of  objects in those extended regions. The first type includes the photospherically detected OB stars. 
Their photospheric properties are determined by applying the spectral classification
criteria as proposed by \citet{Hanson96} and by measuring the shape of
the line profiles. These objects cannot be distinguished
from normal ``naked'' OB stars. This  suggests that these stars have
already reached the main-sequence phase \citep{Hanson02,
Flame03, Ostarspec05}.

In this paper we focus on the spectroscopic properties of the second type of objects;
20 objects that we encountered in our survey of southern UCHIIs. Their spectroscopic and photometric properties differ
from normal main-sequence OB stars and exhibit features commonly
associated with massive young stellar objects. In our
definition, a massive YSO has already contracted to stellar dimensions
and likely commenced with core hydrogen burning, thereby separating a
massive YSO from a high-mass protostar which would still be in its
contraction phase.

 The observational characteristics of massive YSOs are not well
defined, but include: (i) a red continuum, likely due to hot dust,
(ii) a location inside a star forming region and (iii) an emission-line
spectrum (\ion{H}{i}, \ion{Fe}{ii}, \ion{Mg}{ii}, \ion{Na}{i}, CO
first overtone emission). The
detection of emission in the first overtone bands of CO (at
2.3~$\mu$m) is an important signature, as it is considered to be an
indication for the presence of a dense circumstellar disk. CO
first-overtone emission has rarely been detected in any astronomical
object \citep{McGregor88}. The first detection of CO bandhead emission
in a massive YSO was reported by \citet{Scoville79,Scoville83} in the
Becklin-Neugebauer (BN) object. \citet{Geballe87}, \citet{Carr89} and \citet{Chandler93}
detected CO emission in another four high-luminosity YSOs:
\object{S106}, \object{NGC~2024-IRS2}, \object{S140} and
\object{GL2789}. Modelling by \citet{Chandler95} showed that
vibrational temperatures of $\sim 3500$~K are required to produce the
relative band strengths and that the size of the emitting region has
to be quite small ($\sim 1$~AU) and dense ($n \sim
10^{10}$~cm$^{-3}$). A circumstellar disk could provide this
environment, shielding the CO from the hot stellar radiation
field. Alternatively, the CO might be shock-heated due to the
interaction of a stellar wind with the surrounding cloud \citep[see
also][in the case of the \object{BN} object]{Scoville83,Tan04}, or
emerging from the stellar wind itself \citep{Chandler95}. It is not
clear why some YSOs show CO emission and others do not.

\citet{Hanson97} analysed the young
($\la 1$~Myr) stellar population of \object{M17} and uncovered a handful of
massive YSOs. $K$-band spectroscopy revealed that the hottest O stars
are naked, while some of the later O and B stars show clear disk
signatures (near-infrared excess, CO emission, double-lined Pa$\delta$
profiles). Similar conclusions were reached by Blum and
collaborators based on their systematic study of the stellar content
of obscured galactic giant \ion{H}{ii} regions \citep{Blum99,Blum00,
Blum01, Figueredo02, Conti02,Figueredo05}. 

In the following we present $K$-band spectra of 20 candidate massive
YSOs deeply embedded in UCHIIs, obtained with ESO's {\it Very Large
Telescope} and ISAAC.  We study the spectral characteristics of these
potentially very young massive stars, and search for signatures that
may reveal information on the process of their formation.  In the next
section (Sect. \ref{sec:observations}) we introduce our sample and
describe the observations and data reduction process. In
Sect. \ref{sec:spectra} the $K$-band spectra and near-infrared photometry are
presented. In Sect. \ref{sec:refobjects} we discuss some well-studied
objects showing similar signatures as encountered in our massive YSO
sample. In Sect. \ref{sec:discussion} we put the observed properties
of our massive YSO sample in perspective and provide arguments that we
have detected very young massive stars of which at least some are
surrounded by a circumstellar disk, most likely remnant of the
accretion process. We end this paper with a summary of our conclusions
(Sect. \ref{sec:conclusions}).

\section{Sample definition and data reduction process}\label{sec:observations}

The candidate massive YSOs discussed in this paper are selected from a
survey of near-infrared bright point sources deeply embedded in
high-mass star forming regions \citep{Kaper06}. Based on their
position in the colour-magnitude diagram, the most luminous and
reddest sources, and thus the potentially newly born massive stars are
identified. A medium-resolution ($R = 10,000$) $K$-band spectrum has
been obtained with ISAAC on the ESO VLT at Paranal, Chile. This
spectral resolution, as well as the obtained high signal-to-noise
ratio, permits an accurate correction for telluric absorption lines and enables
us to discriminate between stellar and nebular emission. Given the
demand on spectral resolution, a narrow slit ($0.3 \arcsec$) is used,
thus limiting the $K$-band magnitude of the targets to brighter than
$K \simeq 13$.

A number of objects turn out to be late-type foreground stars, but the
majority are of early (OB) spectral type \citep{Ostarspec05}. Many
show a photospheric spectrum for which the $K$-band spectral type has
been determined, resulting in a reliable assessment of their stellar
parameters. Apparently, these stars, though embedded in a young
\ion{H}{ii} region, have already become ``normal'' OB main sequence
stars. Several objects, however, do not show a photospheric spectrum,
but exhibit a broad  Br$\gamma$ emission line; these objects define
the sample of the present paper (Table~\ref{tab:sample}). The
photometric and spectral properties of the 20 candidate massive YSOs
are outlined in the next section.

Long-slit ($120$~\arcsec) $K$-band spectra of these objects are
obtained in two wavelength settings, the first setting with a central
wavelength of 2.134~\micron\ including \brg\ (``\brg-setting''). The
spectrum covered in this setting was used to identify the objects. To
further investigate their potential massive YSO nature, for 15 objects
we also obtained spectra in another setting covering the first four CO
first-overtone bands (central wavelength 2.234~\micron\,
``CO-setting''). The presence of CO first-overtone emission is an
important signature of massive YSOs, and modelling these bands
provides information on the geometry and nature of their circumstellar
material \citep[e.g.][]{COletter04}.

The observations in the \brg-setting were carried out partly in visitor mode (March 18-20, 2000) and partly in service mode (between May 8 and July 16, 2000). The observations in the CO
setting were also obtained in two runs, one visitor mode run (February 21-23, 2002) and a service mode run in the period July 7 to August 2, 2002.

The spectra are electric ghost, dark- and flat-field corrected. The
wavelength calibration of the spectra taken in the \brg\ setting is
performed using arc spectra with the IRAF\footnote{IRAF is distributed
by the National Optical Astronomy Observatories which are operated by
the Association of Universities for Research in Astronomy, Inc., under
cooperative agreement with the National Science Foundation.} task
\emph{identify}. For the spectra obtained in the CO setting the
wavelength calibration is performed using telluric OH emission lines
\citep{Rousselot00}.  The accuracy of the wavelength calibration is
$\sim 3$~\kms. In order to correct for telluric emission the object
was ``nodded'' between two positions on the slit (A and B) such that
the background of position A is recorded while the source is at
position B and is subtracted from the observation where the source is
at position A, and vice-versa. The offset between the A and B positions
was chosen such that the stars\footnote{To increase the efficiency of our
program the orientation of the slit was chosen such that at least two
stars are observed simultaneously.} on the slit do not overlap. In
some objects the nebular emission (in, e.g., the \brg\ line) is
extended so that the emission in the A image is overlapping with the
emission in the B image. After subtraction this results in artifacts
at the position of the nebular lines. We have inspected the 2D spectra
of all sources for the presence of nebular emission and conclude that
in most cases the contribution (positive or negative) of nebular
emission to the stellar (\brg) emission is negligible. 
 However, in a few cases the nebular contribution is significant
  ($\sim50$ \% of the total flux in the case of 06412nr121, 17136nr649 and 18006nr766, see Fig. \ref{fig:kspec1}).

Telluric absorption lines were removed using stars of spectral type A,
observed under identical sky conditions \citep[for the list of
standard stars, we refer to][]{Ostarspec05}.  First, the only
photospheric line (\brg) in the spectrum of the A star needs to be
divided out. It turns out that the best result is achieved when first
the telluric features are removed from the $K$-band spectrum of the
telluric standard using a high-resolution telluric spectrum (obtained
at NSO/Kitt Peak). This spectrum is taken under very different sky
conditions, so a lot of remnants are still visible in the corrected
standard star spectrum, but without this ``first-order'' telluric-line
correction, a proper fit of \brg\ cannot be obtained. The \brg\ line
is fitted by a combination of two exponential functions. The error on
the resulting \brg\ equivalent width (EW) of our target star is about
5~\%. After the removal of the \brg\ line of the A star, the telluric
lines are removed by taking the ratio of the target spectrum with the obtained
telluric spectrum. In the CO setting no stellar features are present
in the telluric standard star spectrum.  The telluric-line correction
is done using the IRAF task \emph{telluric}, which allows for a shift
in wavelength and a scaling in line strength yielding a more accurate
fit.  The task uses a cross-correlation procedure to determine the
optimal shift in wavelength and the scaling factor in line strength,
which can be changed interactively. The shifts are usually a few
tenths of a pixel (1 pixel corresponds to 16 \kms); also the scaling factors are modest ($\sim 10$~\%).

\begin{table*}
%\begin{center}
\begin{tabular}{llllllllllll}\hline\hline
\multicolumn{1}{c}{IRAS}&\multicolumn{1}{c}{H~\sc{ii} region} &\multicolumn{1}{c}{nr}	&\multicolumn{1}{c}{$\alpha$ (2000)}& \multicolumn{1}{c}{$\delta$ (2000)}&  \multicolumn{1}{c}{$K$}& \multicolumn{1}{c}{$J-K$}&\multicolumn{1}{c}{ $A_\mathrm{V}$}&\multicolumn{1}{c}{$K_\mathrm{0}$}  &  \multicolumn{1}{c}{$J-K_\mathrm{0}$}        & \multicolumn{1}{c}{$d$ (kpc)}\\ \hline 
%05393$+$0156&NGC2024   &	IRS2	&05:41:45.7	&-01:54:30.8           	&\phantom{1}5.0  $\pm$ 0.200     & 7.2 $\pm$ 0.22&  28.5 &     -7.0   &       0.9   & 0.4   &  B94                \\ %\hline
\object{06058$+$2138}&\object{AFGL 5180}&	227	&06:08:54.8	&\phantom{-}21:38:49.1 	&10.7 $\pm$ 0.03	        & $>$ 7.0      	&   9 &     -2.0   &\phantom{-}5.5   & 2.2  (K89)              \\ %\hline
\object{06058$+$2138}&\object{AFGL 5180}&	221	&06:08:55.1	&\phantom{-}21:37:55.5 	&10.7 $\pm$ 0.03    	        &3.8 $\pm$ 0.11	&   9 &     -2.0   & \phantom{-}2.3   & 2.2  (K89)              \\ %\hline
\object{06061$+$2151}&\object{AFGL 5182}&        676	&06:09:07.0	&\phantom{-}21:50:31.8 	&13.3 $\pm$ 0.08	        &4.2 $\pm$ 0.50 &  13 &    \phantom{-}0.1   & \phantom{-}2.0   & 2.2  (C95)               \\ %\hline
\object{06084$-$0611}&\object{GGD 14}   &	118	&06:10:50.3	&-06:11:57.6           	&10.8 $\pm$ 0.03	        &4.2 $\pm$ 0.14 &  17 &     -1.1   & \phantom{-}1.3   & 1.0  (K94)               \\ %\hline
\object{06084$-$0611}&\object{GGD 14}   &	114	&06:10:50.3	&-06:11:19.4           	&\phantom{1}9.3  $\pm$ 0.01	&4.7 $\pm$ 0.08 &  17 &     -2.6   & \phantom{-}1.8   & 1.0  (K94)              \\ %\hline
\object{06412$-$0105}&	               &	121	&06:43:48.4	&-01:08:20.6           	&10.2 $\pm$ 0.02	        &1.8 $\pm$ 0.04 &  11 &     -5.3   & \phantom{-}0.0   & 7.3  (Kpc)              \\ %\hline
\object{07299$-$1651}&\object{DG 121}  &	598	&07:32:11.7	&-16:58:32.9           	&11.2 $\pm$ 0.03	        &1.6 $\pm$ 0.06 &  10 &     -0.7   &           -0.1   & 1.4  (W97)                \\ %\hline
\object{07299$-$1651}&\object{DG 121}  &	618	&07:32:09.8	&-16:58:14.7           	&\phantom{1}9.4  $\pm$ 0.01	&4.8 $\pm$ 0.09 &  14 &     -2.9   &  \phantom{-}2.4   & 1.4  (W97)                 \\ %\hline
\object{07299$-$1651}& \object{DG 121}  &	43	&07:32:09.8	&-16:58:14.6           	&\phantom{1}9.4  $\pm$ 0.01	&4.7 $\pm$ 0.09 &  14 &     -2.9   &  \phantom{-}2.3   & 1.4  (W97)                \\ %\hline
\object{07299$-$1651}& \object{DG 121}  &	314	&07:32:09.6	&-16:58:13.8           	&11.5 $\pm$ 0.03	        & $> 6.0$       &  14 &     -0.8   &  \phantom{-}3.6   & 1.4  (W97)                 \\ %\hline
\object{08576$-$4334}&	               &	292	&08:59:21.6	&-43:45:31.6           	&\phantom{1}9.4  $\pm$ 0.01	&2.4 $\pm$ 0.03 &  12 &     -1.2   &  \phantom{-}0.4   & 0.7  (L92)               \\ %\hline
\object{08576$-$4334}&	               &	408	&08:59:28.4	&-43:46:03.6           	&\phantom{1}7.3  $\pm$ 0.01	&3.1 $\pm$ 0.02 &  12 &     -3.3   &    \phantom{-}1.1   & 0.7  (L92)               \\ %\hline
\object{11097$-$6102}&\object{NGC 3576} &	1218	&11:11:53.3	&-61:18:22.0           	&\phantom{1}8.4  $\pm$ 0.01	&5.6 $\pm$ 0.09 &  17 &     -5.8   &    \phantom{-}2.7   & 2.8  (F02)               \\ %\hline
\object{11097$-$6102}&\object{NGC 3576} &	693	&11:11:54.6	&-61:18:23.4           	&\phantom{1}9.2  $\pm$ 0.01	&3.4 $\pm$ 0.04 &  17 &     -4.4   &    \phantom{-}0.5   & 2.8  (F02)               \\ %\hline
\object{15411$-$5352}&	               &	1955	&15:44:59.5	&-54:02:19.2           	&\phantom{1}7.9  $\pm$ 0.01	&4.3 $\pm$ 0.03 &  15 &     -6.0   &    \phantom{-}1.8   & 2.8  (B96)               \\ %\hline
\object{16164$-$5046}&	               &	3636	&16:20:11.3	&-50:53:25.2           	&\phantom{1}9.5  $\pm$ 0.02	&$>$ 8.1       	&  13 &     -4.8   &    \phantom{-}5.9   & 3.6  (K01)              \\ %\hline
\object{16571$-$4029}&\object{RCW 116B} &	1281	&17:00:34.5	&-40:33:40.5           	&\phantom{1}9.3  $\pm$ 0.01	&2.5 $\pm$ 0.06 &  16 &     -2.7   &    \phantom{-}1.5   & 1.0  (C87)               \\ %\hline
\object{17136$-$3617}&\object{GM 1-24}  &	649	&17:17:01.5	&-36:20:57.7           	&\phantom{1}9.5  $\pm$ 0.01	&3.8 $\pm$ 0.06 &  21 &     -4.4   &    \phantom{-}0.3   & 2.0  (T91)               \\ %\hline
\object{17258$-$3637}&	               &	593	&17:29:16.1	&-36:40:07.0           	&\phantom{1}9.6  $\pm$ 0.01	&5.5 $\pm$ 0.17 &  16 &     -4.0   &    \phantom{-}2.8   & 2.3  (B96)              \\ %\hline
\object{18006$-$2422}&\object{M 8}      &	766	&18:03:40.3	&-24:22:39.6           	&\phantom{1}9.4  $\pm$ 0.01	&2.8 $\pm$ 0.04 &  7  &     -2.7   &    \phantom{-}1.7   & 1.9  (C91)              \\ %\hline
\object{18507$+$0110}&\object{AFGL 2271}&	248	&18:53:21.5	&\phantom{-}01:14:00.1 	&10.3 $\pm$ 0.02	        &5.4 $\pm$ 0.18 &  27 &     -5.6   &    \phantom{-}0.9   & 3.7  (W89)               \\ \hline
\end{tabular}
\caption{Sample of candidate massive YSOs. Column 1: IRAS point source
catalogue number; column 2: Name corresponding \ion{H}{ii} region;
column 3: Number $K$-band point source \citep{Kaper06}; columns 4, 5:
Right ascension and declination; columns 6, 7: $K$ and $J-K$
\citep[from][]{Kaper06}; column 8: $A_\mathrm{V}$ applied to deredden
the sources (see text); columns 9, 10: Resulting absolute $K$-band
magnitude and intrinsic $(J-K)_\mathrm{0}$; column 11: Adopted
distance with in between brackets the reference: B96: \citet{Bronfman96} with the galactic rotation
model of \citet{Brand93}, C87: \citet{Caswell87} C91: \citet{Chini81},
C95: \citet{Carpenter95}, F02: \citet{Figueredo02}, K89:
\citet{Koempe89}, K94: \citet{Kurtz94}, K01: \citet{Karnik01}, Kpc:
Kurtz, private communication, L92: \citet{Liseau92}, S90:
\citet{Simpson90}, T91: \citet{Tapai91}, W89: \citet{WoodRadio89}, W97:
\citet{Walsh97}.\label{tab:sample}}
%\end{center}						         
\end{table*}

\section{$K$-band photometric and spectral properties of massive YSO candidates}\label{sec:spectra}

In this section we present the $K$-band spectra of 20 objects selected
from our sample of obscured point sources in high-mass star-forming
regions that display a broad, spectroscopically resolved \brg\
emission line. In Table~\ref{tab:sample} the objects and some of their
observed properties are listed. In the remaining part of the paper we
will name the objects after the first 5 digits (i.e. the right
ascension) of the IRAS point source they are associated with, together
with a number based on our photometry \citep[e.g. object 227 in
\object{IRAS 06058$+$2138} we refer to as \object{06058nr227},
cf.][]{Kaper06}.  These objects are further characterised by a
continuum slope that is on average much redder than that of the OB
stars, that are also present in most of these regions, indicating that these
objects possess an infrared excess. We obtained photometric data in
two near-infrared bands ($J$ and $K$). That is not sufficient to discriminate an IR excess
from extinction in a colour-colour diagram.

In the following we first discuss the photometric properties of these
objects.  After that, we evaluate the spectral features detected in
the \brg\ setting; subsequently, the spectra obtained in the CO
setting are discussed.

\subsection{Position in the colour-magnitude diagram}\label{sec:photom}

\begin{figure*}
\centering \resizebox{\hsize}{!}{\includegraphics{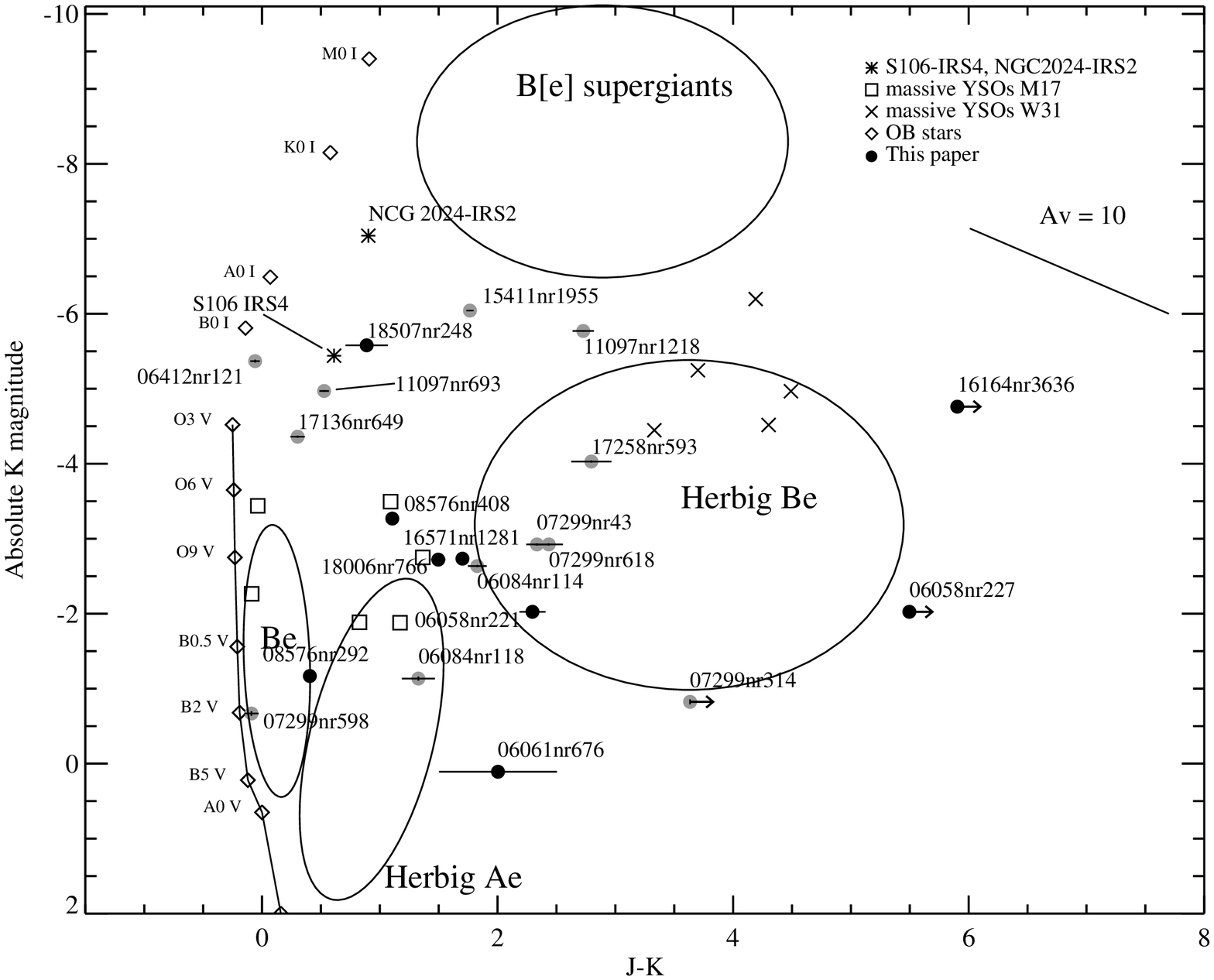}}
\caption{Distribution of massive YSOs in the ($K$,$J-K$) colour
magnitude diagram. The zero-age main sequence (ZAMS) is indicated, as
well as the location of some supergiants ($\diamond$). For each object
an estimate of the amount of interstellar extinction has been obtained
using other members of the embedded cluster. In some cases a spectrum
of a spectroscopically identified OB star \citep{Ostarspec05} could be
used (black circle); alternatively, an average value for the reddening
is derived assuming that the other cluster members are main sequence
stars (grey circle). The diagonal line in the top-right of the figure
is the dereddening line representing 10 magnitudes of visual
extinction. The adopted distance and the uncorrected values of $K$ and
$J-K$ are listed in Table~\ref{tab:sample}. The objects marked by a
``$ \sq $'' and a ``$\times$'' symbol are the extinction corrected
positions of the massive YSOs found in \object{M17} \citep{Hanson97}
and \object{W31} \citep{Blum01}, respectively. The position of
\object{NGC2024-IRS2} \citep{Lenorzer04} and \object{S106}
\citep{Felli84} is indicated by a ``$ \star $''. For comparison, the
areas occupied by Herbig AeBe stars \citep{Eiroa02}, B[e] stars
\citep{Voors99}, and Be stars \citep{Dougherty94} are indicated.
\label{fig:cmd}}
\end{figure*}

In Fig.~\ref{fig:cmd} the objects are plotted in a ($K,J-K$)
colour-magnitude diagram. The observed $K$-band magnitude is
transformed to absolute magnitude by taking into account the distance
and interstellar reddening. The adopted distance is listed in
Table~\ref{tab:sample}. The solid line is the location of the zero-age
main sequence (ZAMS) \citep{Hanson97,Cox00}.

It turns out that practically all candidate massive YSOs are members
of an embedded cluster \citep[cf.][]{Kaper06}. The massive YSOs
located in \object{IRAS 06412-0105} and \object{IRAS 15411-5352} look
isolated, but those regions are associated on a larger scale
with (massive) starforming regions.  Our observations also do not go
deep enough to exclude the presence of low-mass members of the
cluster.  In some cases we have been able to spectroscopically
identify other early-type cluster members \citep{Ostarspec05}. As the
intrinsic colour $(J-K)_{0}$ of all early-type main-sequence stars is
$\sim -0.2$, it is then straightforward to derive the amount of
interstellar reddening.

Assuming that the extinction does not vary a lot over the \ion{H}{ii} region, this method
to estimate the reddening could be applied to nine objects (black
circles in Fig.~\ref{fig:cmd}). For the other objects (grey circles)
the average ($J-K$) of the embedded cluster members was taken to
measure the reddening, assuming that these stars are all on the main
sequence and thus have $(J-K)_{0} \sim 0$. It may well be that the
massive YSOs are more embedded compared to the already emerged OB
stars, and then the derived amount of interstellar extinction is a
lower limit. However, the assumption that the extinction is uniform over the  \ion{H}{ii} region is not likely to be valid. To illustrate this with an example, the visual extinction
measured in the direction of \object{NGC2024-IRS2} is about 3
magnitudes higher than to \object{NGC2024-IRS2b}, just $5 \arcsec$
away \citep{Lenorzer04,Flame03}.

The objects in Fig. \ref{fig:cmd} show a wide range in ($J-K$), 
from 0 to almost 6. Also a number of objects have a very bright $K$
magnitude. If we assume that the $K$-band flux can be fully
attributed to photospheric emission, the absolute $K$-band magnitude of
about half of the stars would be too bright to be main sequence O
stars and would have to be classified as supergiants. The star-forming
regions containing these objects are supposed to be young and unlikely to
host evolved supergiants. This is an indication that the bright $K$-band
magnitude is due to a near-infrared excess.

For most UCHII regions distance determinations are not very
reliable. The distances are mostly based on the radial velocities
measured in molecular lines \citep[e.g.][]{Bronfman96} and converted
to a distance using a rotation model for the galaxy
\citep[e.g.][]{Brand93}. These distance determinations give a near and
a far solution in some directions and, furthermore, the model does not
account for peculiar motion. \citet{Brand93} show that in some
directions the observed radial velocity spread can be up to 30--40
\kms, which makes the distance determination uncertain. If a star of
known spectral type is detected in the embedded cluster, another
constraint on the distance is obtained, so that, e.g., the distance
ambiguity can be resolved \citep[cf.][]{Ostarspec05}.  

For 5  regions a spectroscopic parallax  based on optical data has
been determined. \object{IRAS 17136-3617} and \object{IRAS 18006-2422}
are visible in the optical, while
\object{IRAS 06058$+$2138}, \object{IRAS 06061$+$2151} and \object{IRAS 08576-4334}
are associated with an optically visible \ion{H}{ii} region or OB association (for
references, see column 12 of table \ref{tab:sample}).

\begin{figure*}
\centering \resizebox{\hsize}{!}{\includegraphics{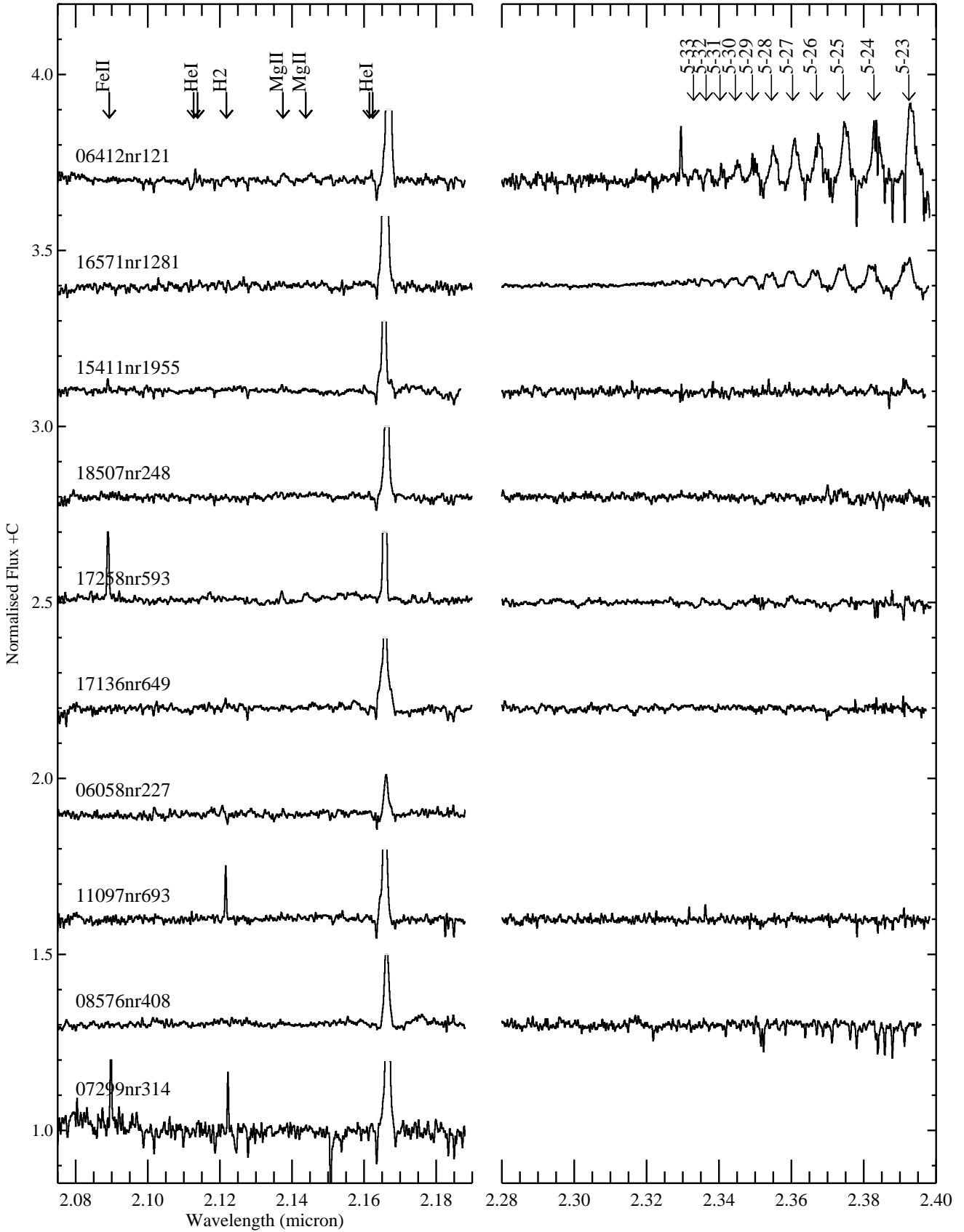}}
   \caption{$K$-band spectra of candidate massive YSOs in two
   wavelength settings: the \brg\ setting (2.08 -- 2.19 \micron,
   \emph{left}) and the CO setting (2.29 -- 2.40 \micron,
   \emph{right}). The spectra are normalised to the continuum. Note
   the gap in wavelength between the two settings. At the top of the
   figure the identification of the spectral lines is given. The CO
   setting displays the most prominent differences between the
   targets. Note the double-peaked Pfund emission lines in
   \object{16571nr1281}. The narrow absorption lines present in the CO
   setting of \object{08576nr408} and \object{17258nr593} are remnant telluric
   lines.  \label{fig:kspec1}}
\end{figure*}

\addtocounter{figure}{-1}
\begin{figure*}
\centering \resizebox{\hsize}{!}{\includegraphics{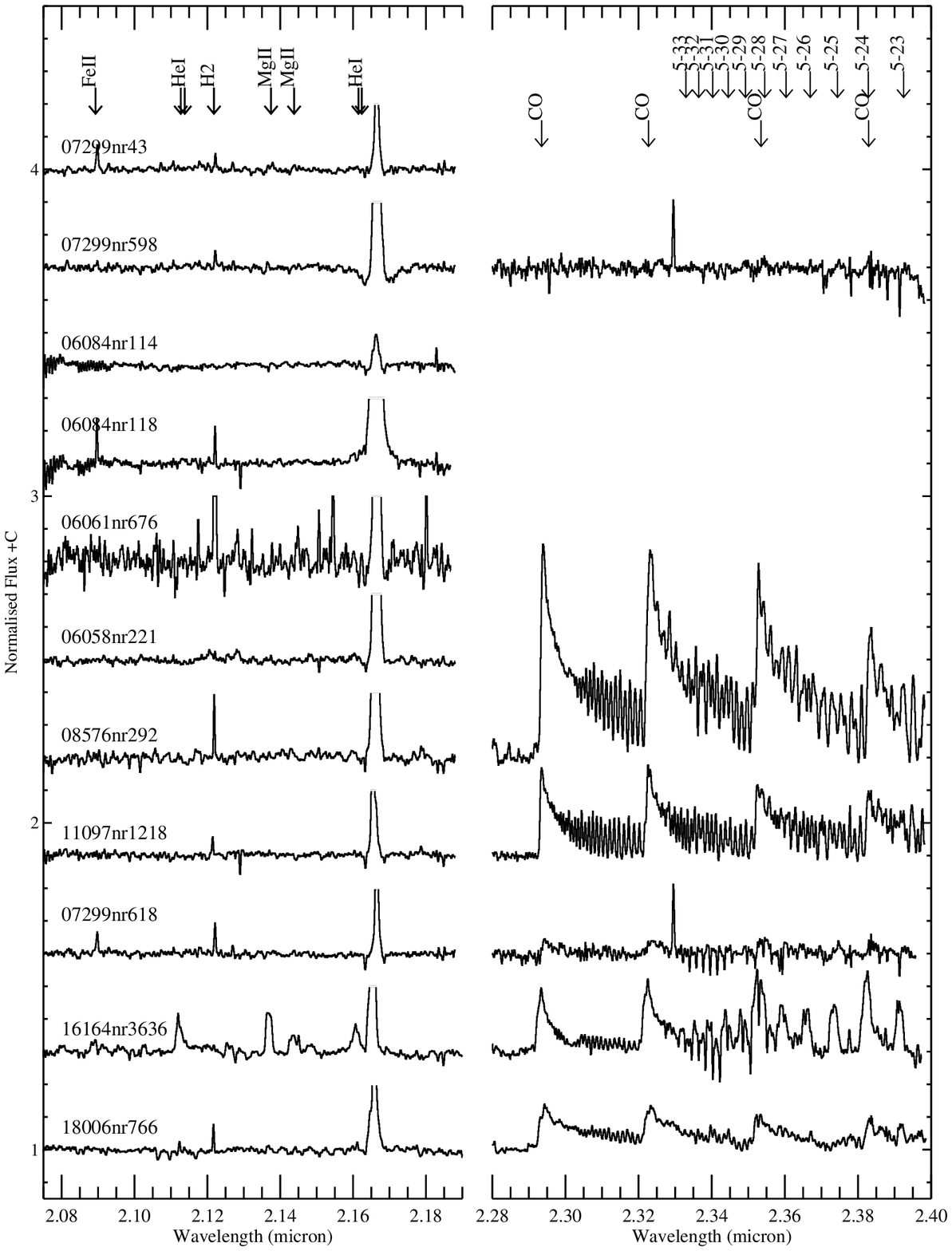}}
   \caption{Normalised $K$-band spectra of candidate massive YSOs
  (continued). CO first-overtone emission is detected in 5 objects
  with pronounced difference in line shape. 16164nr3636 and 18006nr766
  show both CO and Pfund emission lines.  The narrow absorption lines present in
 the CO setting of \object{07299nr598} are remnant telluric lines. \label{fig:kspec2}}
\end{figure*}

\subsection{\brg\ setting\label{sec:brgsetting}}

The objects in our sample were selected based on the presence of a
broad \brg\ emission line. The spectra are displayed in
Fig. \ref{fig:kspec1}. Table~\ref{tab:lineid} lists the
identification of the lines detected in the spectra.  The full
width at half maximum (FWHM) ranges from 100 to 230~\kms\ (the spectral
resolution is $\sim 30$~\kms). The large FWHM suggests that the
emission has a circumstellar (e.g. a rotating disk) and not a nebular origin. In a
\ion{H}{ii} region of 10,000~K the isothermal sound speed is 17 \kms\
and would remain unresolved.  We note, however, that hydrogen
recombination lines detected at radio wavelengths in some
hyper-compact \ion{H}{ii} regions display comparable widths \citep{Kurtz02}.
In two objects (\object{16164nr3636} and \object{06412nr121}) He~{\sc
i} emission lines are detected. In the case of \object{06412nr121}
these lines are not resolved, suggesting a nebular origin. In
\object{16164nr3636} the lines are broad and resolved like the \brg\
line, and probably produced in the circumstellar environment. \hei\
emission is only produced if the nebula is exposed to the strong UV
field of a hot, early-type star \citep[cf.,
e.g.,][]{Hanson02,MartinG2903}. In the majority of objects nebular,
spatially extended, H$_{2}$ 1--0 S(1) emission is observed.  This
emission is produced by molecular hydrogen that is excited by either
shocks or the UV radiation field.

\begin{table}
\begin{center}
\begin{tabular}{lll}\hline\hline
\multicolumn{1}{c}{$\lambda$ (\micron)} &\multicolumn{1}{c}{Element} &\multicolumn{1}{c}{Transition} \\ \hline
\noalign{\smallskip}
  2.0893   &     Fe~{\sc ii} &    3d6.4p-3d6.4s \\
  2.1127   &   \hei  &   1s.3p-1s.4s \\
  2.1138   &   \hei  &   1s.3p-1s.4s  \\
  2.1218   &    \htwo  &    1-0 S(1)\\
  2.1374   &     \ion{Mg}{ii} &   5s-5p \\
  2.1438   &     \ion{Mg}{ii} &   5s-5p \\
  2.1624   &    \hei & 1s.4d-1s.7s \\
  2.1649   &    \hei & 1s.4f-1s.7g \\
  2.1661   &     \ion{H}{i}(\brg) & 7-4  \\
  2.2935   &     CO  &    2-0\\
  2.3227   &     CO &     3-1\\
  2.3535   &     CO &     4-2\\
  2.3829   &     CO &     5-3\\
  2.3329   &     \ion{H}{i}&  33-5\\
  2.3364   &     \ion{H}{i}&  32-5\\
  2.3402   &     \ion{H}{i}&  31-5\\
  2.3445   &     \ion{H}{i}&  30-5\\
  2.3492   &     \ion{H}{i}&  29-5\\
  2.3544   &     \ion{H}{i}&  28-5\\
  2.3603   &     \ion{H}{i}&  27-5\\
  2.3669   &     \ion{H}{i}&  26-5\\
  2.3743   &     \ion{H}{i}&  25-5\\
  2.3828   &     \ion{H}{i}&  24-5\\
  2.3924   &     \ion{H}{i}&  23-5\\\hline
\end{tabular}
\caption{Identification of the lines detected in the $K$-band spectra
  of candidate massive YSOs. Column 1
  lists the rest wavelength, column 2 the corresponding element and
  column 3 the line transition.\label{tab:lineid}}
\end{center}						         
\end{table}

\begin{figure*}
\centering \resizebox{\hsize}{!}{\includegraphics{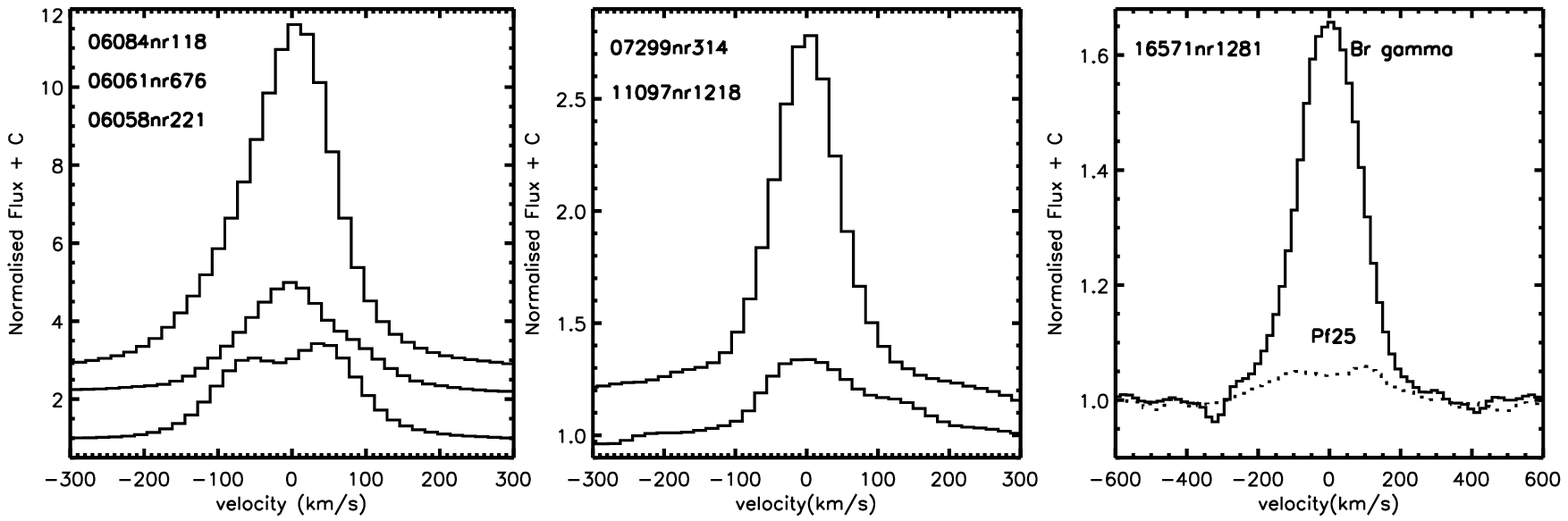}}
\caption{Differences in \brg\ line morphology. The top spectrum in
the left panel  (\object{06084nr118}) shows a blue extended wing, while
the bottom spectrum in the middle panel exhibits a red extended wing of the \brg\
profile. The bottom spectrum of the left panel  (\object{06058nr221}) displays a
double peaked \brg\ line with a peak separation of 95~\kms, suggesting
that it originates in a rotating disk. In the right panel, the \brg\
line and Pf25 line of \object{16571nr1281} are plotted. By measuring the width
of the two lines (Table \ref{tab:pfundlines}), Pf25 turns out to be much broader (and double peaked) than the \brg\
line. Two telluric lines at $-360$ and $345$~\kms\ with respect to \brg\
hamper an accurate determination of the \brg\ line wings. \label{fig:contrast}}
\end{figure*}

The strength of the \brg\ emission covers a wide range (measured
equivalent width (EW) from 1 to over 100~\AA,
with the majority between 1 and 10~\AA, Table~\ref{tab:ew}).  We
do not find a relation between the \brg\ strength and width, nor does
the \brg\ strength correlate with the appearance of other lines
detected in this wavelength setting. One object, \object{06058nr221}, shows a
double-peaked \brg\ profile, with a peak separation of 95~\kms
(Fig. \ref{fig:contrast}). Such a double-peaked profile is also
observed (though in Pa$\delta$) in massive YSOs detected in \object{M17}
\citep{Hanson97}.  The double-peaked nature suggests the presence of a
rotating disk. The \brg\ profile in several other targets displays an
asymmetry. Figure~\ref{fig:contrast} highlights some of the observed
\brg\ lines. Some objects show an additional red component, while
other objects exhibit blue-shifted emission.  There are no signatures
indicating the presence of \ion{C}{iv} and \ion{N}{iii} emission lines,
demonstrating that these stars are not ``normal'' OB supergiants with a
strong stellar wind causing the \brg\ emission \citep{Hanson96}.

Several objects exhibit an \ion{Fe}{ii} emission line (2.089~\micron),
while a few show broad \ion{Mg}{ii} emission (2.138,
2.144~\micron). The \ion{Mg}{ii} emission is likely produced by the
excitation through Ly$\beta$ fluorescence \citep{Bowen47}. Like the
\ion{Fe}{ii} emission, the \ion{Mg}{ii} emission lines indicate the
presence of a dense and warm (several 1000~K) circumstellar environment
\citep{McGregor88}.

\begin{sidewaystable*}
%\centering
%\begin{tabular}{llllllllllll}\hline\hline
\begin{tabular}{p{1.7cm}p{1.1cm}p{1.3cm}p{1.5cm}p{1.3cm}p{1.3cm}p{1.7cm}p{1.3cm}p{1.3cm}p{1.3cm}p{1.3cm}p{1.5cm}}\hline\hline
\multicolumn{1}{c}{Object}&\multicolumn{1}{c}{\ion{Fe}{ii} (2.09)}& \multicolumn{1}{c}{\hei\ (2.1128)}& \multicolumn{1}{c}{\hei\ (2.1137)}& \multicolumn{1}{c}{\ion{Mg}{ii} (2.138)} & \multicolumn{1}{c}{\ion{Mg}{ii} (2.144)}&\multicolumn{1}{c}{ \brg\ (2.166)}   &\multicolumn{1}{c}{FWHM(\brg)}  &\multicolumn{1}{c}{CO}    & \multicolumn{1}{c}{Pf} & &  \\ \hline
\noalign{\smallskip}
\object{06058nr227}  &--	    	    &--	    	   &-- 	    	    &--	     	  &--   	    &\raggedleft1.53$\pm$ 0.10 & \raggedleft 182$\pm$6   & n    &  n   & & \\
\object{06058nr221}  &--	    	    &--	    	   &-- 	    	    &--	     	  &--   	    &\raggedleft36.41$\pm$1.31 &\raggedleft 202$\pm$1   & n    &  n  & & \\
\object{06061nr676}  &--	    	    &--	    	   &-- 	    	    &--	     	  &--   	    &\raggedleft33.22$\pm$ 0.34&\raggedleft 166$\pm$11  & n    &  n   & & \\
\object{06084nr118}  &0.60$\pm$0.04          &--	    	   &-- 	    	    &--	    	  &--   	    &\raggedleft111.1$\pm$ 0.12&\raggedleft 156$\pm$1   & n    &  n  & &  \\
\object{06084nr114}  &--	    	    &--	    	   &-- 	    	    &--	     	  &--   	    &\raggedleft1.51$\pm$ 0.11 &\raggedleft 212$\pm$8   & n    &  n   & & \\
\object{06412nr121}  & --	            &0.13$\pm$0.03 & 0.07$\pm$0.03  &0.78$\pm$0.18&0.76$\pm$0.17    &\raggedleft12.2$\pm$ 0.08 & \raggedleft147$\pm$1   & --   &  +   & &\\
\object{07299nr598}  &0.24$\pm$0.04         &--	    	   &-- 	    	    &-- 	  &--   	    &\raggedleft19.6$\pm$ 0.13 &\raggedleft 217$\pm$1   & --   &  -- & & \\
\object{07299nr618}  &0.48$\pm$0.04         &--	    	   &-- 	    	    &-- 	  &--   	    &\raggedleft4.12$\pm$ 0.06 &\raggedleft 138$\pm$2   & +    &  -- & & \\
\object{07299nr43}   &0.53$\pm$0.05         &--	    	   &-- 	    	    &0.28$\pm$0.11&0.25\raggedleft$\pm$0.16    &\raggedleft4.87$\pm$ 0.08 &\raggedleft 146$\pm$1   & --   &  -- & & \\
\object{07299nr314}  &1.43$\pm$0.06         &--	    	   &-- 	    	    &-- 	  &--  	    	    &\raggedleft14.9$\pm$ 0.11 &\raggedleft 115$\pm$2   & --   &  -- & &  \\
\object{08576nr292}  &--	    	    &--	    	   &-- 	    	    &-- 	  &--  	    	    &\raggedleft15.1$\pm$ 0.16 & \raggedleft220$\pm$2   & +    &  -- & & \\
\object{08576nr408}  & --	            &--	    	   &-- 	    	    &--     	  &--  	    	    &\raggedleft3.88$\pm$ 0.09 &\raggedleft 181$\pm$2   & --   &  -- & & \\
\object{11097nr1218} &--	    	    &--	    	   &-- 	    	    &-- 	  &--  	    	    &\raggedleft4.83$\pm$ 0.08 &\raggedleft 191$\pm$3   & +    &  -- & & \\
\object{11097nr693}  &--	    	    &--	    	   &-- 	    	    &-- 	  &--  	    	    &\raggedleft6.73$\pm$ 0.07 &\raggedleft 121$\pm$1   & --   &  -- & & \\
\object{15411nr1955} &0.22$\pm$0.04         &--	    	   &-- 	    	    &0.47$\pm$0.10&0.34$\pm$0.10    &\raggedleft6.99$\pm$ 0.06 &\raggedleft 106$\pm$1   & --   &  +  & & \\
\object{16164nr3636} &--	    	    &2.15$\pm$0.22 &-- 	    	    &2.17$\pm$0.15&1.25$\pm$0.24    &\raggedleft16.66$\pm$ 0.10&\raggedleft 132$\pm$1   & +    &  +  & & \\
\object{16571nr1281} &--	    	    &--	    	   &-- 	    	    &--	     	  &--  	    	    &\raggedleft10.65$\pm$ 0.13&\raggedleft 204$\pm$1   & --   &  +  & & \\
\object{17136nr649}  &--	    	    &--	    	   &-- 	    	    &--	     	  &--  	            &\raggedleft6.06$\pm$ 0.10 &\raggedleft 182$\pm$2   & --   &  -- & & \\
\object{17258nr593}  &1.61$\pm$0.04         &--	    	   &-- 	    	    &0.33$\pm$0.10&0.67$\pm$0.20    &\raggedleft7.98$\pm$ 0.06 &\raggedleft 118$\pm$2   & --   & --  & & \\
\object{18006nr766}  &--	    	    &0.13$\pm$0.03 &0.06$\pm$0.04   &--	     	  &--  	            &\raggedleft10.65$\pm$ 0.08&  \raggedleft 99$\pm$1    & +    &  +  & & \\
\object{18507nr248}  &--	    	    &--	    	   &-- 	    	    &--	     	  &--  	   	    &\raggedleft6.41$\pm$ 0.07 &\raggedleft 154$\pm$1   & --   &  -- & & \\ \hline\hline
\multicolumn{1}{c}{Object}& \multicolumn{1}{c}{Pf23}& \multicolumn{1}{c}{Pf24}& \multicolumn{1}{c}{Pf25}& \multicolumn{1}{c}{Pf26}& \multicolumn{1}{c}{Pf27}& \multicolumn{1}{c}{Pf28}& \multicolumn{1}{c}{Pf29}& \multicolumn{1}{c}{Pf30}& \multicolumn{1}{c}{Pf31}& \multicolumn{1}{c}{Pf32}& \multicolumn{1}{c}{Pf33} \\ \hline
\noalign{\smallskip}
\object{06412nr121}  & 4.45$\pm$0.23 & 3.27$\pm$0.28& 3.76 $\pm$0.26 &2.61$\pm$0.27&2.58$\pm$0.26    & 2.00$\pm$0.25 &1.04$\pm$0.23&0.94$\pm$0.21 & 0.50$\pm$0.11 & 0.66$\pm$0.18  & 0.67$\pm$ 0.25 \\
\object{154111995}   & 0.53$\pm$0.01 & 0.41$\pm$0.01& 0.43 $\pm$0.01 &--           &--               & --            & --          & --           & --            &  --            & --             \\
\object{16164nr336}  & 3.15$\pm$0.17 & b            & 3.46 $\pm$0.19 &3.20$\pm$0.20& 3.35$\pm$0.21   & b             & 1.85$\pm$0.19& 2.00$\pm$0.18& --           & --             & --             \\
\object{16571nr1281} & 2.81$\pm$0.10 & 2.29$\pm$0.10& 2.01$\pm$0.10  &1.33$\pm$0.10&1.34$\pm$0.10    & 1.02$\pm$0.10 & 0.88$\pm$0.09& 0.53$\pm$0.08& 0.47$\pm$0.13& 0.52$\pm$0.15  & 0.55$\pm$0.23 \\
\object{18006nr766}  & 1.06$\pm$0.06 & --           & --             &--           &--               & --            & --          & --           & --            &  --            & --             \\ \hline \hline
\multicolumn{1}{c}{Object}      & \multicolumn{1}{c}{CO(2--0)}       & \multicolumn{1}{c}{CO(3--1)}      & \multicolumn{1}{c}{CO(4--2)}        & \multicolumn{1}{c}{CO (5--3)}    &                 &               &             &              &               &                &                \\ \hline
\noalign{\smallskip}
\object{07299nr618}  &\raggedleft 4.0          &\raggedleft b            & \raggedleft   b           &\raggedleft 1.0        &                 &               &             &              &               &                &                \\ 
\object{08576nr292}  &\raggedleft 67.1        &\raggedleft 76.5        &\raggedleft    64.5       &\raggedleft  32.3       &                 &               &             &              &               &                &                \\ 
\object{11097nr1218} &\raggedleft 25.0         &\raggedleft  28.7       & \raggedleft     28.7      &\raggedleft  19.1      &                 &               &             &              &               &                &                \\ 
\object{16164nr3636} &\raggedleft 12.4          &\raggedleft b            &\raggedleft     b          &\raggedleft  b          &                 &               &             &              &               &                &                \\ 
\object{18006nr766}  &\raggedleft 17.2         &\raggedleft 16.2        &\raggedleft 12.2          & \raggedleft b          &                 &               &             &              &               &                &                \\ \hline
\end{tabular}
      \caption{Equivalent width (EW) measurements of the emission line
	objects. \emph{Upper table:} Columns 2--7 give the EW for the
	spectral lines in the \brg\ setting. The EWs are measured in \AA
	ngstr\"{o}m and emission is given as positive. In column 8 the
	FWHM of \brg\ is given in \kms. Columns 9 and 10 show whether
	the CO and Pfund-lines are present in the spectra or
	not. (``--'' means not present, ``n'' means not
	observed). \emph{Middle table:} The EW measurements of the
	Pf-lines. The ``b'' means that the line could not be measured
	because of blending with the CO lines.  The Pf-lines of
	\object{16571nr1281} are double peaked with an average peak
	separation of $189 \pm 33$~\kms. \emph{Lower table:} The EW
	measurements of the CO first-overtone bandheads. The ``b''
	means that the lines are blended with cold CO absorption lines
	or Pfund lines.\label{tab:ew}}
\end{sidewaystable*}

\subsection{CO setting}\label{sec:cosetting}

\subsubsection{CO first-overtone emission}\label{sec:cobandheads}

One of the most remarkable spectral features found in these objects is
the CO first-overtone emission in the range from $2.29-2.40$~\micron.
CO bandheads are detected in 5 sources (Fig.~\ref{fig:kspec2}), with
different shapes of the bandhead. The objects \object{11097nr1218} and
\object{08576nr292} show a very steep bandhead, while the other 3
objects have a more shallow bandhead shape and exhibit an extended
blue wing. The shape of the bandhead reflects the velocity dispersion
of the CO molecules. In \object{11097nr1218} and \object{08576nr292}
the CO gas has a small velocity dispersion, which also results in a
high contrast in the rotational $J$ lines. In the other objects the CO
molecules have a much larger velocity dispersion, resulting in a
reduced contrast of the $J$ lines and a broadening of the
bandheads.

The observed extended blue wing of the bandhead cannot be
explained by a gaussian velocity distribution. A keplerian velocity
distribution, however, can reproduce such a bandhead shape.
\citet{Chandler93,Chandler95} have observed and modelled high
resolution CO 2--0 observations of a sample of YSOs and modelled the
kinematics of the CO gas.  These models are based on a circumstellar
accretion disk or a neutral stellar- or disk wind. Most of the objects
display a CO bandhead with a blue shoulder, for which
\citet{Chandler95} obtained the best fit with a circumstellar disk.  A
detailed analysis and modelling of the CO first-overtone bands in our
massive YSO sample support this conclusion and is presented in \citet{COletter04}. Similar results have been obtained by  \citet{Blum04}. 
Contrary to the large difference in shape of the CO bandheads, the
FWHM of  \brg\  is rather constant, also the CO lines with a
small velocity dispersion show a broad FWHM of \brg\ (Table
\ref{tab:ew}). This suggests that \brg\ is not formed in the
same region/geometry as the CO bandheads.

In the spectra of \object{16164nr3636} and \object{07299nr618}, also
CO absorption lines are detected ($2.33 - 2.35$~\micron). These lines
are the lower $J$-lines of the 2--0 bandhead and indicate cold ($\sim
40$~K), foreground molecular gas.%({\bf REF})

\subsubsection{Pfund lines}\label{sec:pfund}

In five objects hydrogen Pfund lines are detected, including two which
also have the CO first-overtone bands in emission. In the spectrum of
16571nr1281 the Pf-lines are double peaked with a separation of $189
\pm 33$~\kms\ suggestive of a disk origin. In Table~\ref{tab:pfundlines}
the FWHM of Pf25 is given in comparison with the \brg\ line. It turns
out that the Pf25 line is always much broader than \brg,
which would be consistent with a disk origin (Sect. 5). 

The Pf-line emission indicates a line forming region of high density
($N_\mathrm{e} \sim 10^{8} \mathrm{cm}^{-3}$).  The flux ratio of Pf25 and \brg\ is listed in the last column of Table~\ref{tab:pfundlines}.
The ratio varies from 0.08 to 0.3, in all cases larger than what is expected in the optically thin case B scenario \citep[$<$ 0.02,][]{Storey95}. This means that when the emission of the Pf25 and \brg\ are emitted by the same gas, this gas must be  (partially) optically thick. However, when the two lines are produced at different locations, this ratio provides us with the ratio of the two line forming surface areas assuming that the emitting material is fully optically thick.

If the Pf-lines
originate in a keplerian disk, they are likely formed in the inner
disk region, while for example \brg\ is formed in a much more extended
part of the disk. In such a scenario the FWHMs of the Pf-lines and
\brg\ are expected to be different: \brg\ is predominantly formed in
the more slowly rotating outer parts of the disk (larger surface area)
and thus will have a smaller FWHM.

\begin{table}
\begin{center}
\begin{tabular}{llll}\hline\hline

\multicolumn{1}{c}{Object}&\multicolumn{1}{c}{FWHM \brg} & \multicolumn{1}{c}{FWHM Pf25}  & Pf25/\brg\\ 
            & \multicolumn{1}{c}{(\kms)}          & \multicolumn{1}{c}{(\kms)}    & \\ \hline
\noalign{\smallskip}
\object{08576nr292}   & 220 $\pm$ 2     			& -- 						& --	 	\\
\object{11097nr1218} & 191 $\pm$ 3     			& -- 						& --		\\
\object{07299nr618}   & 138 $\pm$ 2     			& --  						& --		\\
\object{16164nr3636} & 132 $\pm$ 1        			& 273 $\pm$ \phantom{3}8 	& 0.3		\\
\object{18006nr766}   & \phantom{1}99  $\pm$ 1     	& {\it 282 $\pm$ 13} 			& 0.1		\\
\object{06412nr121}   & 147 $\pm$ 1                          	& 269 $\pm$ \phantom{3}7 	& 0.2 	\\
\object{15411nr1955} & 106 $\pm$ 1                          	& 140 $\pm$ 37 			& 0.08		\\
\object{16571nr1281} & 204 $\pm$ 1                          	& 328 $\pm$ \phantom{3}8 	& 0.2		\\ \hline
\end{tabular}
      \caption{The FWHM of the \brg\ line compared with that of the
      hydrogen Pf25 line. The Pf23 line is stronger than the Pf25, but
      this line is heavily contaminated by residuals of atmospheric
      lines. For the fitting of \object{18006nr766} the Pf23 line is used,
      because Pf25 is not detected. In the last column the flux-ratio of \brg\ and Pf25 is listed. \label{tab:pfundlines}}
\end{center}
\end{table}

\section{Comparison to well studied objects}\label{sec:refobjects}

In order to better understand the nature of the objects presented in this
paper as well as the physical conditions under which the spectral
lines are formed, the observed photometric and spectral properties of the
candidate massive YSOs are compared to those of well-studied objects
in the same area of the CMD.

In the following section we will first discuss some 
objects of which the massive YSO nature has been established based on
a broad range of observations. After that we discuss the supposedly
``more evolved'' objects like Be and B[e] stars which show similar
$K$-band spectra.
 In the last part of this section the observed
properties of the candidate massive YSOs are discussed in detail and
compared to those of the reference objects. 
 The existing spectra of those objects are usually of worse quality and resolution than the spectra of the candidate YSOs presented in this paper. This makes a quantitative comparison difficult.

 When objects have similar near-infrared spectra, this does not necessarily mean that they are in the same
phase of stellar evolution; it does, however, indicate that both
classes of objects provide the physical conditions under which a
certain near-infrared spectrum is produced.

In Table \ref{tab:refobjects} the near-infrared photometric
and spectroscopic properties of the reference objects discussed in this section are summarized. Below, a short description of the special characteristics of the object or classes of objects is given.

\begin{table*}[!ht]
\begin {center}
\begin{tabular}{lllllllccr} \hline \hline
\multicolumn{1}{c}{Object/}	& \multicolumn{1}{c}{Sp. Type}	& \multicolumn{1}{c}{\brg}	&\multicolumn{1}{c}{Pf}	& \multicolumn{1}{c}{HeI} 	&  \multicolumn{1}{c}{CO}	&  \multicolumn{1}{c}{MgII/FeII}	&\multicolumn{1}{c}{\emph{E(J-K)}}	&\multicolumn{1}{c}{\emph{K}}	&\multicolumn{1}{c}{References}\\
\multicolumn{1}{c}{Object type}&						&					&					&					&					&						&						& 					&					\\ \hline
S106					& O6--O8 V				&+					&+					&--					& +					& -- 						&0.6						&-5.4				&G87, C89, C93, C95, vdA00	\\
BN						& B0.5V					&+					&--					&--					& + 					&+						&--						&--					&S83, G87, C89			\\
NGC2024-IRS2			&B0V					&+					&+					&--					& +					&+						& 0.9						&-7.0				&G87, C93, C95, L04	 	\\
Herbig Ae					& F - late B				&+					&--					&--					&--					&--						& 1.5						& 2 -- -2				&E02, vdA04 			\\
Herbig Be					& B						&+ 					&?					&+					&some				&--						& 2 -- 6					&-1 -- -5				&E02, vdA04 		\\
Be						& B						&+					&+					&some				&--	 				&some					& 0.0 - 0.6					& 0 -- -3				& D94, C00			 \\	
B[e]						& B						&+					&+					&some				&+					&+						& 2 -- 6$^\dagger$			&-6 -- -10$^\dagger$	&M88,M96,V99			\\\hline
\end{tabular}
\caption{Summary of the spectral and photometric properties of the well studied objects as discussed in Sect. \ref{sec:refobjects}. Column 1: Object name; column 2: Spectral type; columns 3-6: Spectral features detected in their K-band spectra; columns 7 \emph{(J-K)} value, dereddened for foreground extinction; column 8: Absolute, dereddened \emph{K}-band magnitude; column 9: References:  G87: \citet{Geballe87}, C89: \citet{Carr89}, C93: \citet{Chandler93}, C95: \citet{Chandler95}, vdA00: \citet{vandenAncker00}, S83: \citet{Scoville83}, L04: \citet{Lenorzer04}, vdA04: E02: \citet{Eiroa02}, \citet{vandenAncker04}, C00: \citet{Clark00}, D94: \citet{Dougherty94}, M88: \citet{McGregor88}, M96: \citet{Morris96}, V99: \citet{Voors99}. $\dagger$: These values are for the B[e] supergaints.}\label{tab:refobjects}
\end{center}
\end{table*}

\subsection{(Massive) Young Stellar Objects}\label{sec:refyso}

\noindent
\emph{\object{S106-IRS4}} \\ \object{S106} is one of the prototype
massive YSOs. The broad shape of the CO bandheads in the K-band spectrum (Table \ref{tab:refobjects}) is explained by emission from a narrow region in a circumstellar disk \citep{Chandler95}.
The source is further characterised by a bi-polar outflow ($v_\mathrm{exp}$
= 200 \kms), with a very pronounced equatorial gap.  The gap is also
visible in the radio, where extinction does not play a role
\citep{Bally83}. This  indicates that the gap is not caused by additional
extinction, but is probably the result of a shadow produced by a
circumstellar disk, which is supporting the results of the fits of the CO bandheads.

\smallskip

\noindent
\emph{BN object}\\ Among the YSOs, the Becklin-Neugebauer (BN) object
is one of the best studied. 
% \citet{Scoville83} find
%that the central star produces a UV radiation field equivalent to that
%of a B0.5~V star ionising an \ion{H}{ii} region.
The CO bandheads are rather steep. If
the CO emission would come from a disk, the modest velocity broadening
leads to the conclusion that the disk is viewed almost
pole-on. \citet{Scoville83} suggest that CO emission comes from a
shock-heated thin layer far from the star caused by the motion of the
BN object through a cold medium.

\smallskip

\noindent
\emph{\object{NGC2024-IRS2}}\\ The infrared source
\object{NGC2024-IRS2} was discovered by \citet{Grasdalen74} and
long thought to be the ionising source of the Flame Nebula
(\object{NGC2024}). However, \citet{Flame03} identified
\object{NGC2024-IRS2b} as the ionising source, a late O, early B
star.  Using simple gas and
dust models of prescribed geometry \citet{Lenorzer04} argue that the
infrared flux from the circumstellar material of \object{NGC2024-IRS2}
is produced by a dust-free, dense gaseous disk.  The density in the disk of \object{NGC2024-IRS2} is on
the order of $10^{14} - 10^{15} \mathrm{cm}^{-3}$ in the equatorial
plane at the stellar surface. This is 2--3 orders of magnitude higher
than in classical Be stars \citep{Waters91}.

\smallskip
\noindent
\emph{Massive YSOs in (giant) \ion{H}{ii} regions}

A near-infrared study of the young \ion{H}{ii} region \object{M17} by
\citet{Hanson97} has revealed a sample of massive stars possessing a
near-infrared excess. The $K$-band spectral properties are identical
to the objects presented here.  Spectra obtained around 1 micron revealed double peaked \pad\ emission,
superposed on photospheric absorption lines. In the optical where
the circumstellar material does not produce excess emission, the
photospheric spectrum of the underlying star is visible. The spectral
types derived from these spectra are early and mid-B. 
\smallskip

\noindent
\emph{Herbig Ae/Be stars}\\ 
The circumstellar emission of Herbig Ae stars is produced by a passive, dusty
circumstellar disk \citep[e.g.][]{Mannings97,Millan-Gabet01,Eisner03}. 
For the Herbig Be stars the geometry of the circumstellar
material is not obvious. At sub-mm wavelengths the predicted emission
produced by a cold, massive circumstellar disk is not detected,
suggesting that the geometry of the circumstellar material of the
Herbig Be stars is different from that of the Herbig Ae
stars. \citet{Natta00} suggest that this may reflect a difference in
time scales. The dispersal of the in-falling envelope by radiation
pressure may trigger a rapid evolution of the circumstellar disk, and
the disk will disappear much faster than in the case of Herbig Ae
stars. When the Herbig Be stars become optically visible, their disks
may already have disappeared. \citet{Fuente03} found
evidence for a disk around 3 Herbig Be stars, but the derived masses
($10^{-3} - 10^{-2} ~\rm{M}_{\sun}$) are substantially lower than the
mass of the disks around Herbig Ae stars ($10^{-2} - 10^{-1}~
\rm{M}_{\sun}$).

\subsection{Main-sequence and more evolved stars}\label{sec:refevolved}

\emph{Be stars}\\ Be stars are rapidly rotating main sequence or
(sub)giant stars that are characterised by H$\alpha$ emission
produced by high density circumstellar gas. Often the H$\alpha$ line
is double peaked.  Their infrared spectra are dominated by hydrogen
lines \citep{Hony00,Vandenbussche02}, including many from higher
energy levels.   Interferometric
observations have confirmed the hypothesis that the circumstellar gas
is distributed in a disk-like geometry \citep{Vakili98,Stee98}. The
physical mechanism producing and maintaining these circumstellar disks
is not known, but is likely related to the rapid rotation of the central
star.

\smallskip

\noindent
\emph{B[e] stars}\\ 
 B[e] stars show  hydrogen lines in
emission but also include forbidden lines in their optical spectra,
mainly of [Fe\ {\sc ii}] and other low ionisation species.  Contrary to Be stars, these
stars exhibit a strong near-infrared and mid-infrared excess. The SED
peaks between 5 and 10~\micron, showing the signature of a hot (500 --
1000~K) dust component \citep{Voors99}.  The B[e] stars comprise stars
of very different nature. The classification scheme proposed by
\citet{Lamers98} includes a class of supergiants as well as a class of
pre-main sequence objects. However, for many B[e] stars the evolutionary status
is unknown. The infrared excess of B[e] supergiants is likely caused by a highly non-spherical wind which is much denser and slower at the equator than in the polar regions \citep{Zickgraf85}. This is likely due to the rapid rotation
of the star \citep{Pelupessy00}

\smallskip
In Fig.~\ref{fig:cmd} the objects discussed above are plotted,
together with the ZAMS and the massive YSOs presented in this
paper. The Be stars are located close to the main sequence. The Herbig
Be stars are much redder and bridge the gap between the Herbig Ae
stars and the B[e] supergiants. The majority of the candidate massive
YSOs are found near the location of the Herbig Be stars, suggesting that
these objects are also of B spectral type. A small group
of objects show a brighter $K$-band magnitude and also a bluer
($J-K$) colour. They are located near the well-known massive YSO
\object{S106-IRS4}. Also \object{NGC2024-IRS2} is found in that area
of the CMD. The location of this group suggests that the central stars
of these objects are of O spectral type, i.e. hotter and more
luminous than the Herbig Be stars. For some objects, this suggestion
is supported by additional arguments (see below).

%\subsection{Discussion of the different  classes in the CMD}\label{sec:objects} 
\subsection{Classification of our objects}\label{sec:objects}

\begin{table*}
\begin{tabular}{lllllllllp{4.5cm}}
\multicolumn{1}{c}{(1)} & \multicolumn{1}{c}{(2)}&\multicolumn{1}{c}{(3)} & \multicolumn{1}{c}{(4)}&\multicolumn{1}{c}{(5)}&\multicolumn{1}{c}{(6)}    &\multicolumn{1}{c}{(7)} &\multicolumn{1}{c}{(8)}    &\multicolumn{1}{c}{(9)}    &\multicolumn{1}{c}{(10)}\\\hline\hline
\multicolumn{1}{c}{Object} & \multicolumn{1}{c}{CO}&\multicolumn{1}{c}{\ion{Mg}{II}/} & \multicolumn{1}{c}{Pf}&\multicolumn{1}{c}{UCHII}          &\multicolumn{1}{c}{Disk}    &\multicolumn{1}{c}{near-IR} &\multicolumn{1}{c}{IRAS}    &\multicolumn{1}{c}{Other}    &\multicolumn{1}{c}{Cluster}\\
                           &                       &\multicolumn{1}{c}{\ion{Fe}{II}}     & \multicolumn{1}{c}{}     &\multicolumn{1}{c}{} &\multicolumn{1}{c}{}& \multicolumn{1}{c}{Sp. type}&\multicolumn{1}{c}{Sp. type}&\multicolumn{1}{c}{Sp. type} &\multicolumn{1}{c}{Properties}\\ \hline
\object{06058nr227} &  n   & $-$  & n    & $-$ &    & B?      & B1V   &               &Cluster member, B1-B2V star found in same cluster. \\ 
\object{06058nr221} &  n   & $-$  & n    & $-$ & +  & late B  & B1V   &               &See \object{06058nr227}. \\ 
\object{06061nr676} &  n   & $-$  & n    & +   &    & B?      & B1V   & B2V (K94)     &Cluster member, B1-B2V star found in same cluster, central source UCHII region.\\   
\object{06084nr118} &  n   & +    & n    & +   &    & mid-B   & B1.5V & $>$B3 (P03)   &Cluster member (\object{GGD 14}), negative radio slope (G02). Red mid-IR SED (P03). \\ 
\object{06084nr114} &  n   & $-$  & n    & $-$ &    & mid-B   & B1.5V &               &Cluster member.\\ 
\object{06412nr121} &  $-$ & +    & +    & +   & +  & mid-O   & O9.5V & O8V (Kpc)     &Nebular \hei\ emission, ionising\\
                    &      &      &      &     &    &         &       & O7V (N84)     & source UCHII; brightest source. \\  
\object{07299nr598} &  $-$ & $-$  & $-$  & $-$ &    & B2V     & B2V   & late B (C00)  &Likely not related to IRAS source, Be star. \\ 
\object{07299nr618} &  +   & $-$  & $-$  & $-$ & +  & mid-B   & B2V   &               &Scattered light of more embedded source? \\ 
\object{07299nr43}  &  +   & +    & $-$  & $-$ &    & mid-B   & B2V   &               &Blended with \object{07299nr43}. \\ 
\object{07299nr314} &  $-$ & +    & $-$  & $-$ &    & B?      & B2V   &               &Deeply embedded. \\ 
\object{08576nr292} &  +   & $-$  & $-$  & $-$ & +  & mid-B   & B1V   &               &Located at edge of the cluster, 2 O9V-B1V stars in center. \\ 
\object{08576nr408} &  $-$ & $-$  & $-$  & $-$ &    & early-B & B1V   &               &See \object{08576nr292}. \\ 
\object{11097nr1218}&  +   & $-$  & $-$  & $-$ & +  & early-B & O6V   &               &Located in cluster, not main ionising source, peak of radio emission even more embedded. \\ 
\object{11097nr693} &  $-$ & $-$  & $-$  & $-$ &    & late-O  & O6V   &               &Blue mid-IR SED (B03), located close to \object{11097nr1218}. \\ 
\object{15411nr1955}&  $-$ & +    & +    & n   & +  & mid-O   & O8V   &               &Nebular \hei\ emission, main ionising source \ion{H}{ii} region. No radio observations available. \\ 
\object{16164nr3636}&  +   & +    & +    & n   & +  & mid-O   & O6V   & $>$O4V (C87)  &Main ionising source very compact \ion{H}{ii} region (3.5\arcsec); nebular \hei\ emission, embedded in massive envelope (K01). \\ 
\object{16571nr1281}&  $-$ & $-$  & +    & $-$ & +  & early-B & B0.5V & $>$O9V (C87)  &Cluster member, next to O8V-B1V star. \\ 
\object{17136nr649} &  $-$ & $-$  & $-$  & $-$ &    & late-O  & O8V   &               &In center of small cluster (GM24), next to UCHII region \\ 
\object{17258nr593} &  $-$ & +    & $-$  & $-$ &    & early-B & O8V   &               & \\ 
\object{18006nr766} &  +   & $-$  & +    & $-$ & +  & mid-B   & O9.5V &               &Located in M8, next to the ionising source \object{Her 36} (O7V).  \\ 
\object{18507nr248} &  $-$ & $-$  & $-$  & $-$ &   & late-O  & O5.5V &                &Located in cluster, also O5-O6 V star detected \\ \hline
\end{tabular}
\caption{Summary of the properties of the massive YSOs. Columns 2--4:
  Spectral lines ($-$: not present, $+$: present, $n$: not observed); column 5: Is the source the counterpart of a radio
  UCHII?; column 6: Spectral evidence for a
  circumstellar disk. ($+$: evidence for a disk); column 7: Spectral type estimated based on the
  position in the CMD (Fig. \ref{fig:cmd}); column 8: Spectral type
  based on the IRAS luminosity of the IRAS source. This is an upper
  limit on the spectral type of the massive YSO, as usually more stars
  contribute to the IRAS flux. Column 9: Spectral type estimate based
  on other methods. Column 10: Properties of the surrounding cluster,
  taken from \citet{Kaper06}. The OB star spectra are discussed in
  \citet{Ostarspec05}. References: K94: \citet[ radio flux UCHII
  region]{Kurtz94}; P03: \citet[ mid-infrared luminosity]{Persi03};
  Kpc: Kurtz, private communication (radio flux UCHII region); N84:
  \citet[ optical spectroscopy]{Neckel84}; C00: \citet[ \brg\
  absorption, similarity with Be stars]{Clark00}; C87: \citet[ single
  dish radio observations and assuming that a single star is
  responsible for the ionisation]{Caswell87}; G02: \citet{Gomez02};
  B03: \citet{Barbosa03}; K01: \citet{Karnik01}. \label{tab:summary}}
\end{table*}

A detailed description of the fields and stellar populations hosting
the massive YSO candidates discussed in this paper is given in
\citet{Kaper06}. Here we focus on the characteristics relevant for a
better understanding of the physical nature of these objects. In Table
\ref{tab:summary} a summary is given of the properties of the
objects. The objects will be discussed in groups based on their position in the
CMD (Fig. \ref{fig:cmd}).

\subsubsection{Objects with B spectral type}\label{sec:Btypes}

The majority of the candidate massive YSOs have positions in the CMD
which suggest that they are of B spectral type. A few objects hardly
show any infrared excess and have colours similar to Be stars
(\object{07299nr598} and \object{08576nr292}, Fig. \ref{fig:cmd}). The
spectral types of these objects (Table \ref{tab:summary}, column 7)
are estimated by subtracting a typical $K$-band excess for Be stars ($\sim$0.6
magnitude).  The $K$-band spectrum of \object{07299nr598} also has the
characteristics of a Be star. Underneath the \brg\ emission line, a
broad photospheric absorption component is detected. This is also seen
in the late-type Be stars \citep{Clark00}. The spectrum of
\object{08576nr292}, however, does not have the spectral characteristics of
Be stars and shows strong CO bandheads. 

The other objects with a larger $J-K$ are located in the same area as
the Herbig Be stars. None of the stars show similarities
with the Herbig Ae stars, indicating that the stars we have detected are
more massive. The spectral properties, however, are less
homogeneous. Some objects show CO bandheads, some Pf-lines, while
other objects only show \brg\ emission. This difference in spectral
signature is, however, also seen in Herbig Be stars.

\subsubsection{Objects with O spectral type \label{sec:Otypes}}

A small group (5 objects) is located close to \object{S106} in the
CMD. Their luminosity suggests that the central stars are  O
stars.  They have a relatively blue colour ($J-K \sim 0-2$), suggesting a
similar structure of the circumstellar material
(Sect. \ref{sec:destruction}). Mid-infrared data of
\object{11097nr693} indicate a very blue SED \citep{Barbosa03}, likely
dominated by circumstellar gas instead of dust.

The spectral properties, however, are not similar to \object{S106},
but also differ among each other.  Apart from the \hei\ emission, the
spectral features do not depend strongly on the luminosity of the
stars, but are much more determined by the physical conditions, like
temperature and density in the circumstellar material. The main effect
which seems to depend on the luminosity is the relatively blue
colour of the more massive objects.%{\bf ?}

 The object which resembles the spectroscopic properties of
\object{S106} most is \object{16164nr3636}. The presence of cold
CO absorption lines and the position in the CMD suggest that this
object is deeply embedded. The extinction of the neighbouring OB
stars  used to deredden \object{16164nr3636} is obviously much
less than the real extinction in this line of sight. The
CO-bandheads in this object posses a blue wing which is naturally
explained by a keplerian rotating disk \citep{COletter04}.

\subsubsection{Objects showing strong \brg\ emission}\label{sec:brgemission}

In the majority of the
candidate massive YSOs in our sample the \brg\ EW is around 10~\AA\ or
less. In three objects, however, the EW is significantly higher:
\object{06058nr221}, \object{06061nr676}, \object{06084nr118} (30--100~\AA,
Table \ref{tab:ew}). These objects are all located in the lower part of
the CMD (Fig.\ref{fig:cmd}).  They are of equal luminosity as the
Herbig Ae stars, but significantly redder ($J-K \sim 2$).

Mid-infrared observations show that \object{06084nr118} has a very
red mid-infrared SED \citep{Persi03}, suggesting that the source is
still heavily embedded. The IR-luminosity they derive is 350
L$_{\sun}$, which corresponds to an embedded star with a spectral type
later than B3.  The radio spectral energy
distribution has a negative slope, is variable and has a non-thermal
origin. \citet{Gomez02} suggest that the radio emission is originating
in an active magnetosphere, like observed towards T-Tauri stars. The
derived IR-luminosity as well as the position in our CMD suggest that
this object is much more luminous than a T-Tauri star. It has a
$K$-band magnitude similar to Herbig-Ae stars.

The \brg\ line of \object{06058nr221} is double peaked with a
separation of 95~\kms, suggesting an origin in a rotating disk. The
disk hypothesis has not been confirmed by CO or Pfund lines, as
no observations were performed in this setting. \object{06061nr676} is
one of the few objects being the near-infrared counterpart to a radio
UCHII region (see Sect. \ref{sec:ionising}), which might be
contributing to the \brg\ emission.

\subsection{Ionising sources of UCHII regions}\label{sec:ionising}

\begin{figure*}
\centering \resizebox{\hsize}{!}{\includegraphics{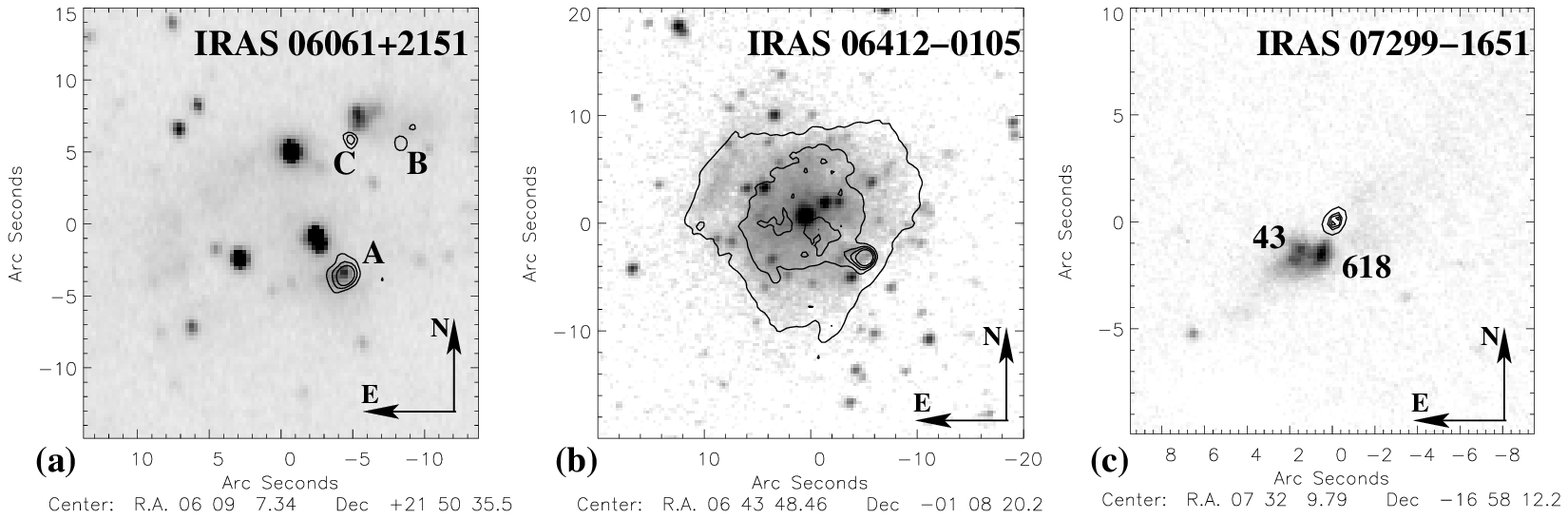}}
\caption{$K$-band images overlayed with radio continuum data of 3 UCHII
  regions. Panel ({\bf a}): $K$-band image of \object{IRAS 06061+2151}
  \citep{Kaper06} overlayed with a VLA 3.6~cm image \citep{Kurtz94}. A
  (\object{G188.796+01.030}), B (\object{G188.793+01.030}) and C (\object{G188.794+01.031})
  correspond to 3 UCHII regions, with UCHII region A coinciding with
  the \object{06061nr676}. Panel ({\bf b}) $K$-band image of \object{IRAS 06412-0105}
  \citep{Kaper06} overlayed with a VLA 1.3~cm image (Kurtz,
  priv. comm.). Panel ({\bf c}) VLT $K$-band acquisition image of \object{IRAS
  07299-1651}, where the position of the radio source is indicated
  \citep{Walsh98}. None of the three sources are coinciding with the
  radio source, which may be even more embedded.  \label{fig:radiodata}}
\end{figure*}

\citet{Kaper06} find a near-infrared counterpart of an UCHII radio
source in 44 \% of the fields containing such a source. This is in
agreement with the results of \citet{Hanson02}, who show that only
half of the radio UCHII regions have a near infrared counterpart.

The UCHII regions without a near-infrared counterpart are apparently still so
heavily embedded that they are not detectable at near-infrared wavelengths. By comparing the positions
of the candidate massive YSOs with those of the UCHII regions, we find
that 3 massive YSOs in our sample are coincident with a radio source
(see  Table \ref{tab:summary}, column 5), although the radio
counterpart of one of them, 06084nr118, does not have the radio
properties of a typical UCHII region (see
Sect. \ref{sec:brgemission}).

To illustrate this, in panel (a) and (b) of Fig. \ref{fig:radiodata} two
sources are displayed with a radio counterpart, while panel (c) shows an
example of a region where the radio source is not the counterpart of
the massive YSO, but of a more embedded source.  In panel (a), the
$K$-band image of \object{IRAS 06061+2151} demonstrates that
\object{06061nr676} is part of a larger cluster and the only source
coincident with a radio source (source A: \object{G188.796+01.030}).
Panel (b) shows the $K$-band image of \object{IRAS 06412-0105} where
\object{06412nr121} falls in the center of the radio emission. Because
of the large distance to this source (7.3 kpc), the extent of the
region (0.8 pc) suggests that this region is not an UCHII but
a more evolved \ion{H}{ii} region.

\section{Discussion}\label{sec:discussion}

We have observed a large sample of candidate massive YSOs located in
high-mass star-forming regions.  The observed photometric and spectral
characteristics show that these objects are surrounded by
circumstellar material. The next question to address is whether this
material is located in a circumstellar disk and, if this is the case,
whether these disks  are remnants of the accretion process, or formed
as a consequence of the mass-loss process and rapid rotation such as
in Be and B[e] stars.

\subsection{Origin of the spectral lines}\label{sec:lines}

%{\bf Add sketch indicating line-forming regions}.

%\begin{figure*}
%\centering \resizebox{\hsize}{!}{\includegraphics{disk_clear_colors.eps}}
%\caption{}
%\end{figure*}

The different spectral lines observed in the $K$-band spectra of the
candidate massive YSOs are formed in regions with different physical
conditions. These conditions provide information on  the origin of
the material. In some of the objects the spectral lines indicate that
they are formed in a rotating disk (see Table
\ref{tab:summary}, column 6).

\subsubsection{High-density indicators}

The special conditions required to emit the CO first-overtone emission
are suggestive of an origin in a circumstellar disk. The CO is only
emitted at high density because the critical density of the CO
overtone transitions is high ($10^{10} - 10^{11}~\rm{cm}^{-3}$). The
excitation temperature has to be high as well ($2000-4000~\rm{K}$).
The disk hypothesis is supported by the shape of the CO bandheads of two
objects (\object{16164nr3636} and \object{18006nr766}). They show a CO
bandhead profile displaying a blue wing, which can be explained by
 CO gas in keplerian rotation \citep{COletter04}.

CO bandheads can also be produced in a shock front. This is suggested
for the BN object where \citet{Scoville83} propose that the CO is
formed in the shock-front between the \ion{H}{ii} region and the
ambient molecular cloud. Another tracer of shocks is \htwo, which is
also observed towards our candidate massive YSOs. However, the \htwo\
is spatially extended, and possibly formed in the photo-dissociation
region by UV excitation. The CO is only observed towards the point
sources, suggesting that the CO and \htwo\ emission are not formed by the same
process.

Pfund-lines probe high-density material which is also an indication for a
disk-like geometry.  This is confirmed for the Pf-lines in
\object{16571nr1281}, which are double peaked, indicating that the
lines are emitted in a disk. The Pf-lines in \object{16164nr3636} and
\object{18006nr766} have a width comparable to that of the CO
bandheads. This suggests that the lines share the same kinematics. They
cannot be formed in the same region, as the Pf-lines are formed in
ionised gas with a  temperature too high to coexist with CO molecules. This
leads to the suggestion that, in case of a circumstellar disk, the CO
is formed in the neutral mid-plane, and the Pf-lines originate in the
ionised upper layers. 

Two of our reference objects, \object{S106} and \object{NGC2024-IRS2},
show similar bandhead profiles. In these objects  also other evidence
for a circumstellar disk is found. In the case  of \object{S106},
evidence for a disk comes from the observed geometry of the extended optical
and near-infrared emission \citep[cf.][]{Smith01}. For \object{NGC
  2024-IRS2}, the infrared SED points to a (gaseous) disk \citep{Lenorzer04}.

\subsubsection{Origin of \brg\ emission}

The majority of the objects, however, do not show Pfund or CO
emission and therefore might have another geometry of the
circumstellar material. The \brg\ line, however, which is present in
all the spectra, shows a remarkably similar behaviour in all
objects. This might suggest that the formation mechanism of the \brg\ line
could be similar for all objects.  Especially the FWHM of \brg\ is
rather constant, it varies between 100 and 220~\kms.  The EW of \brg\
ranges from 1 to 10~\AA, except in 3 objects where the EW is much
larger (upto 100~\AA, see Table~\ref{tab:ew}). Similar EW values are
found in Be stars, where the larger EWs are found for the objects
with the earlier spectral type (B0--B2) \citep{Clark00}.

If the circumstellar matter is, like the CO and Pf-lines suggest, in
the shape of a circumstellar disk,  \brg\ emission may be produced  in
the ionised upper layers of the disk. The \brg\ emission will be
formed over a fairly large area of the circumstellar disk and the
emission will be dominated by the largest surface area. This in
contrast to the Pf-lines which are only formed in the high-density
inner parts of a disk, which makes them much broader than the \brg\
line (Table \ref{tab:pfundlines}).

The width of the \brg\ line suggests that the line is not formed in the
\ion{H}{ii} region itself. The sound speed in (ultra) compact
\ion{H}{ii} regions is too low to account for the  \brg\ line widths. For
an isothermal \ion{H}{ii} region of 10\,000 K, the sound speed is
about 17~\kms. We note that in hyper-compact \ion{H}{ii} regions broad
radio recombination lines up to 180~\kms\ are observed
\citep{Kurtz02}, which probably reflects their expansion
velocity. These objects are heavily embedded, much more than the UCHII
regions and will, very likely, not be visible in the near-infrared.

 The large width of the \brg\ line could also be explained by a bipolar outflow. 
The red-shifted and the blue-shifted lobes tilted in the line of sight would then be responsible for the observed width. However, the other spectral lines which then also should be produced in these outflows have different FWHMs, some are broader and some are even unresolved. This would be unlikely in the case of a bipolar outflow.  Additionally, our objects show a very strong continuum which is not observed in that case.

Also a normal stellar wind as observed for OB stars would not
explain the \brg\ emission lines. Only in supergiants the \brg\ line
is in emission. But these stellar winds have terminal velocities
of the order of 1500~\kms. Supergiants are not expected in young
star-forming regions.  If the lines are formed in a stellar wind,
\brg\ would be broader than the Pf-lines \citep{Lamers96}, which is
not seen in the data (Table \ref{tab:pfundlines}).

It is also possible that the \brg\ line is formed in a disk-wind
\citep{Drew98}. If the central star is emitting a lot of UV radiation,
it will ionise the surface of the disk so that the disk 
evaporates (see Sect. \ref{sec:destruction}). This will then result
in a dense and low velocity disk wind, where the outflow velocity will
soon dominate the rotational velocity. The line widths predicted
by \citet{Drew98} are of the order of 200~\kms.  This scenario would
also explain the difference in width of the CO and \brg\ lines in
\object{08576nr292} and \object{11097nr1218}, where the CO can be
formed in a disk seen face on and \brg\ is originating from a
disk-wind.

Summarising, the broad CO emission as well as the double-peaked
Pf-lines strongly suggest that the material in these objects is
located in a circumstellar disk. By comparing the behaviour of \brg\
in all the objects, the origin in a disk-wind seems the most plausible
explanation. This means that all the objects may be surrounded by a
circumstellar disk. The absence of CO bandheads in the spectra of most
of the objects might therefore be related to the special conditions
required to emit the first-overtone emission, rather than the
abundance of the CO molecules.

\subsection{Nature of the objects}\label{sec:nature}

The aim of this project is to identify the ionising source(s) of UCHII
regions, and to determine their physical nature. The regions were
selected based on the detection of an UCHII radio source and its
connection with an IRAS point source fulfilling the UCHII
colour-colour criterion of \citet{WoodIRAS89}. The near-infrared
surveys of UCHII regions performed by \citet{Hanson02} and
\citet{Kaper06} demonstrate that only half of the UCHII radio sources
show a near-infrared counterpart.  After comparison of the positions
of the candidate massive YSOs with those of the UCHII radio positions, only a
few objects turn out to be the actual near-infrared counterpart of an
ultra-compact radio source (see Sect. \ref{sec:objects}). All objects,
however, are part of more extended sites of (massive) star formation
which also host UCHII regions.  

%Apparently, often the ionising source(s) of the UCHII
%regions are still too much embedded to be detectable in the
%near-infrared. 
%Probably only the more evolved UCHII regions show a
%near-infrared counterpart, while the counterparts of the really young
%UCHII regions cannot be observed in the near-infrared.

The fact that the candidate massive YSOs are located inside regions
of star formation suggests that these objects are young. However, also
around main sequence OB stars, disks are observed (Be and B[e] stars).
The densities derived for the disks of Be stars are too low to emit
the CO bandheads \citep[$\sim 10^{10}
\rm{cm}^{-3}$,][]{Waters91}. Also disk-winds are not observed in Be
stars. Studies of Be star abundances in open clusters suggest that
the Be phenomenom mainly occurs in the second half of their main
sequence lifetime \citep{Fabregat00}.  In B[e] stars, CO emission is
observed; \citet{Kraus03} show that in a very dense, non-spherical
stellar wind, such densities can be reached. Only supergiants have
these strong stellar winds.  None of the candidate massive YSOs is
found near the location of the B[e] supergiants in the CMD
(Fig. \ref{fig:cmd}).

The arguments presented above suggest that the majority of these
objects are young  and their circumstellar disks are likely a
product of the formation process of these massive stars. These disks
might be the remnant of the extended accretion disks detected in
earlier phases of massive star formation
\citep[e.g.][]{Shepherd01Science,Beltran04,Garay99,Chini04,Jiang05,Patel05}.

\subsection{Disruption of disks around young massive stars}\label{sec:destruction}

In some of the  star-forming regions where the massive YSOs are
detected, also massive stars are found which already cleared their
circumstellar environment. Observations of \object{M17}
\citep{Hanson97} as well as giant \ion{H}{ii} regions \citep{Blum03}
give similar results.

In our spectroscopic sample we find that most of the O stars ``already''
cleared out their environment while a relatively larger fraction of
the B stars are still surrounded by circumstellar material. This leads
to the suggestion that the most massive stars clear out their
environment fastest.

This trend is also seen when inspecting the location of the massive YSOs
in the CMD.  The majority of the massive YSOs have a position in the
CMD indicating spectral type B.  A few objects, however, show evidence
that the central star is a hot O star (see \S \ref{sec:objects}). The
location of these sources in the CMD is near \object{S106} for which a
spectral type O6--O8 has been proposed \citep[e.g.][]{vandenAncker00}.
Note that the position of \object{NGC2024-IRS2} also suggests an O
star as the central source. However, when taking into account a strong
$K$-band excess and the identification of the main ionising source of
\object{NGC2024}, \object{NGC2024-IRS2b} (O8), it is more likely that
\object{NGC2024-IRS2} is an early B star \citep{Flame03,Lenorzer04}.

The objects where the central star likely is an O star show a ($J-K$)
colour which is on average much bluer than those of the objects which
are of spectral type B. This difference in colour suggests that the
disk properties of these objects are different. O stars produce much
more EUV radiation and possess stronger stellar winds than B stars,
which shortens the destruction timescale of disks around O stars.

\citet{Hollenbach94,Hollenbach00} investigated the different
destruction mechanisms of disks around young stars. For OB stars, the
outer regions of the disk will be photoevaporated. The extreme-UV
photons ionise the top-layer of the disk. This gas will be heated to
$10^4$~K and obtain a velocity dispersion given by the local sound speed
($\sim 10$~\kms). Near the star this velocity will be smaller than the
escape velocity. The gravitational radius ($r_\mathrm{g}$), the radius
at which the sound-speed is equal to the escape velocity ranges from
$\sim 500$~AU for O3 stars to $\sim 100$~AU for early B
(Tab.~\ref{tab:diskdestruct}). Beyond the gravitational radius, the
sound speed will exceed the escape velocity so that the matter can
freely escape from the disk, e.g. in the form of a disk wind.

This model discriminates between two scenarios, the weak wind case and
the strong wind case. For massive stars, the strong-wind case is
applicable. In this strong-wind case, the disk is flattened by the
stellar wind.  The strong stellar wind prohibits that ionised gas
located beyond $r_\mathrm{g}$ is freely escaping from the disk surface, but
is dragged over the surface of the disk to the point where the ram
pressure of the wind balances the thermal pressure of the ionised
hydrogen. From there on, the gas can freely escape from the disk. This
strong-wind case leads to a faster destruction of the disk than the
weak-wind case.

The timescale as derived by \citet{Hollenbach94}
depends on the number of ionising continuum photons, the mass-loss
rate, the wind velocity and the mass of the disk. The stellar
parameters are taken from \citet{Mokiem04} and \citet{Lenorzer04}.  By
assuming a disk mass of 2 solar masses
\citep[see][]{Hollenbach94,Yorke96}, the destruction timescale ranges
from somewhat less than $10^{5}$ year for a disk around an early O star,
to $\sim 5\times 10^{5}$~year for a disk surrounding a B0~V star
(Table~\ref{tab:diskdestruct}). The assumption in this model is that
the material in the disk is not replenished, and that accretion has
stopped.  These short timescales indicate that already very soon after
the accretion stops, the outer disk regions will be rapidly destroyed by
the EUV photons of the star.

For the destruction of the regions located within the gravitational
radius, the direct interaction with the stellar wind is much more
important. By considering the momentum transferred by the collision of
a spherical wind with the disk \citet{Hollenbach00} derive a timescale
for this process. Typical timescales for the destruction of the inner
parts of the disk by the interaction with the stellar wind are given
in column 6 of Table~\ref{tab:diskdestruct}. The mass of the disk
inside this gravitational radius is taken to be 10\% of the total disk
mass.

\begin{table}
\begin{center}
%\begin{tabular}{p{0.6cm}p{0.75cm}p{1.3cm}p{0.75cm}p{0.75cm}p{0.9cm}p{0.9cm}}\hline\hline
\begin{tabular}{llllll} \hline \hline
\multicolumn{1}{c}{SpTp} &\multicolumn{1}{c}{ log($Q{_\mathrm{0}}$)} &   \multicolumn{1}{c}{log(\mdot)}& \multicolumn{1}{c}{$r_\mathrm{g}$}     & \multicolumn{1}{c}{$t_\mathrm{p}$}  &  \multicolumn{1}{c}{$t_\mathrm{w}$} \\  
   &\multicolumn{1}{c}{ s$^{-1}$}     & \multicolumn{1}{c}{M$_{\sun}$~yr$^{-1}$} & \multicolumn{1}{c}{AU}    & \multicolumn{1}{c}{10$^{5}$~yr}   & \multicolumn{1}{c}{10$^{5}$~yr} \\ \hline
\noalign{\smallskip}
O3V &        49.69 &       -5.375 &              459.5 &        0.6 &        0.7\\
O4V &        49.50 &       -5.599 &              377.4 &        0.8 &        1.1\\
O5V &        49.31 &       -5.805 &              301.4 &        0.9 &        1.8\\
O5.5V &      49.08 &       -6.072 &              251.4 &        1.1 &        3.3\\
O6.5V &      48.82 &       -6.369 &              205.4 &        1.3 &        6.5\\
O7.5V &      48.51 &       -6.674 &              169.4 &        1.7 &        13.3\\
O9V &        48.06 &       -7.038 &              141.4 &        2.5 &        30.7\\
O9.5V &      47.79 &       -7.252 &              128.7 &        3.3 &        50.1\\
O9.7V &      47.52 &       -7.445 &              118.0 &        4.2 &        78.8\\
B1V &        46.92 &       -7.913 &               98.7 &        7.3 &        233.5\\\hline
\end{tabular}
      \caption{Two different effects are responsible for the
      destruction of circumstellar disks around O stars; the
      photo-evaporation process and the direct interaction between the
      wind and the disk. In the first 3 columns, the stellar
      parameters are given used in the calculation of the timescales
      \citep{Mokiem04,Lenorzer04}. In column 4, the gravitational
      radius is given. Beyond this radius matter is removed by
      photoevaporation, while inside of it, the stellar wind directly removes
      material from the disk. Col. 5 and 6 give the typical
      timescales for these two processes. The formulae are taken from
      \citet{Hollenbach94,Hollenbach00}.\label{tab:diskdestruct}}
\end{center}
\end{table}

Comparison of the two timescales (Table~\ref{tab:diskdestruct}) shows
that first the outer layers of the disk are destroyed by
photo-evaporation and that the inner parts of the disk are removed at
a later stage. So observationally one should look for indicators of
a remnant disk by  probing material near the star, i.e. warm gas and
dust, and not the cold material probed at mm wavelengths.

The detected number of massive YSOs which harbour an O star compared
to the amount of massive YSOs which are of spectral type B is
qualitatively 
consistent with the trend suggested by this model.  The destruction of
the disks around young O stars is much faster than those around B
stars, so the probability to detect an O star surrounded by a remnant
accretion disk is much smaller than such a disk around a young B star. This
suggests that these objects are really young massive stars surrounded
by a remnant accretion disk. Observations of Herbig Be stars also show
the effect of the outer parts of the disk being photoevaporated by the
UV photons. The disk masses, if detected at all, are substantially
lower than those around Herbig Ae stars \citep{Natta00,Fuente03}.

\section{Conclusions}\label{sec:conclusions}

In this paper, high-resolution $K$-band spectra of candidate massive
YSOs located in high-mass star-forming regions are presented. The
results can be summarised as follows:

\begin{itemize}
\item[$\bullet$]{The near-infrared photometric properties of the
  massive YSO candidates suggest the presence of
  a near-infrared excess. This implies that large amounts of
  circumstellar matter are present around these objects.}
\item[$\bullet$]{This prediction is confirmed by VLT/ISAAC $K$-band
  spectroscopy. The objects were selected on the presence of \brg\
  emission, but also show other emission lines formed in a warm
  circumstellar environment like \ion{Fe}{ii}, \ion{Mg}{ii}, Pf 23-5
  -- Pf 33-5 and CO
  first-overtone bandhead emission. In some objects also \hei\ emission is detected.}
\item[$\bullet$]{In general these candidate massive YSOs are not the ionising
  sources of UCHII radio sources. Only 3 sources are associated
  with an UCHII radio source. All the objects, however, are located
  in more extended \ion{H}{ii} regions; in most cases also OB
  stars are found which already cleared out their environment. The
  UCHII radio sources located in these fields are not visible in the
  near-infrared, likely because they are deeply embedded.}
\item[$\bullet$]{The location of these objects inside high-mass star-forming
  regions, with associated UCHII radio sources, suggests that
  these objects are still in an early evolutionary phase, although the
  objects may be more evolved than the ionising sources of the UCHII
  regions which often remain undetected. The massive YSOs found in \object{M17}
  and a sample of giant \ion{H}{ii} regions suggest that these objects have an
  age of about 10$^6$~year.}
\item[$\bullet$]{Most of the stars have near-infrared colours similar to the
  Herbig  Be stars, suggesting that also most of the stars in our
  sample are of spectral type B. For 5 objects, however, the location
  in the CMD (Fig. \ref{fig:cmd}) suggests that the central stars are
  of spectral type O. For two sources this is supported by
  circumstantial evidence provided by the nebular emission lines and
  IRAS infrared flux.}
\item[$\bullet$]{Photo-evaporation and disk destruction models presented by
  \citet{Hollenbach94,Hollenbach00} predict that O stars clear their
  circumstellar environment very soon after the accretion has stopped. For
  B stars the timescales are substantially longer. This is consistent
  with the detection of only a few  massive YSOs with spectral
  type O.}
\item[$\bullet$]{The \brg\ line likely originates in an ionised disk wind,
  while the Pf lines are probably emitted in the inner
  regions of a disk. The CO emission lines are only produced in a very dense
  medium, and are likely forming in the dense mid plane of the
  disk.}
\item[$\bullet$]{The high density to emit the CO bandheads as well
  as the origin of \brg\ in a disk-wind do not point to  an
  origin in a mass-loss disk such as present around Be
  stars. These conditions are met in the dense non-spherical winds of
  B[e] supergiants, but the location of the sources in high-mass
  star-forming regions as well as their location in the CMD
  (Fig. \ref{fig:cmd}) are not consistent with a possible supergiant nature.}
\item[$\bullet$]{The evidence presented in this paper suggests that
  the infrared excess is created by a remnant accretion disk, rather
  than a mass-loss disk. These disks are likely the remnants of the
  large accretion disks found around high mass stars in hot cores
  \citep{Minier98,Shepherd01Science, Beltran04,Chini04}.}
\end{itemize}

\begin{acknowledgements}
AB acknowledges financial support from the DFG during a two-month
visit at ESO Headquaters. LK acknowledges the support of a fellowship
of the Royal Academy of Arts and Sciences in the Netherlands. The
authors thank the VLT staff for support and help with the
observations, Margaret Hanson for help with the observations and
datareduction process, Annique Lenorzer, Wing-Fai Thi, Stan Kurtz and
Fernando Comer\'on for helpful discussions.  Andrew Walsh and Ed
Churchwell are thanked for providing the radio images. We thank the referee for his/her suggestions to improve the paper. NSO/Kitt Peak
FTS data used here were produced by NSF/NOAO.
\end{acknowledgements}

%\bibliographystyle{aa}
%\bibliography{../arjan}
\end{document}